%% file: SMP-22-011_temp.tex
\pdfoutput=1
\documentclass[11pt,twoside,a4paper,cmspaper,final,collab]{cms-tdr}
\def\svnVersion{8b9538f}\def\svnDate{2026/03/17}\def\cmsCernNoTag{CERN-EP-2025-147}\def\cmsCernDate{\today}\def\cmsMessage{Published in the Journal of High Energy Physics as \href{http://dx.doi.org/10.1007/JHEP03(2026)022}{\doi{10.1007/JHEP03(2026)022}.}}
\begin{document}\cmsNoteHeader{SMP-22-011}

\newcommand{\alpEW}{\ensuremath{\alpha_\text{EW}}}
\newlength\cmsTabSkip\setlength{\cmsTabSkip}{1ex}
\ifthenelse{\boolean{cms@external}}{\providecommand{\cmsTable}[1]{\resizebox{0.49\textwidth}{!}{#1}}}{\providecommand{\cmsTable}[1]{#1}} 

\cmsNoteHeader{SMP-22-011}
\title{Vector boson scattering and anomalous quartic couplings in final states with \texorpdfstring{$\Pell\PGn\PQq\PQq$}{lnuqq} or \texorpdfstring{$\Pell\Pell\PQq\PQq$}{llqq} plus jets using proton-proton collisions at \texorpdfstring{$\sqrt{s} =$ 13\TeV}{sqrt(s) = 13 TeV}}

\abstract{A measurement is presented of the electroweak vector boson scattering production of $\PZ\PV$ ($\PV = \PW, \, \PZ$) boson pairs associated with two jets in proton-proton collisions at a center-of-mass energy of 13\TeV. The data, corresponding to an integrated luminosity of 138\fbinv, were collected at the CERN LHC with the CMS detector during the 2016--2018 data-taking period. The analysis targets final states with a pair of isolated electrons or muons from \PZ boson decays and three or four jets, depending on the momentum of the vector boson that decays into quarks. Signal strength is measured for events characterized by a large invariant mass of  two forward jets with a wide pseudorapidity gap between them. The electroweak production of $\PZ\PV$ in association with two jets is measured with an observed (expected) significance of 1.3 (1.8) standard deviations. A combination of the analyses of $\PZ\PV$ channel and the previously published $\PW\PV$ channel in the lepton plus jets final state places constraints on effective field theory parameters that describe anomalous electroweak production of $\PW\PW$, $\PW\PZ$, and $\PZ\PZ$ boson pairs in association with two jets. Several world best limits are set on anomalous quartic gauge couplings in terms of dimension-8 standard model effective field theory operators.}

\hypersetup{
pdfauthor={CMS Collaboration},
pdftitle={Vector boson scattering and anomalous quartic couplings in final states with \texorpdfstring{$\Pell\PGn\PQq\PQq$}{lnuqq} or \texorpdfstring{$\Pell\Pell\PQq\PQq$}{llqq} plus jets using proton-proton collisions at \texorpdfstring{$\sqrt{s} =$ 13\TeV}{sqrt(s) = 13 TeV}},
pdfsubject={CMS},
pdfkeywords={CMS, SMP, VBS, EFT, semileptonic, Machine Learning}}

\maketitle 

\section{Introduction}

\begin{figure}[htb]
    \centering
    \includegraphics[width=0.49\textwidth]{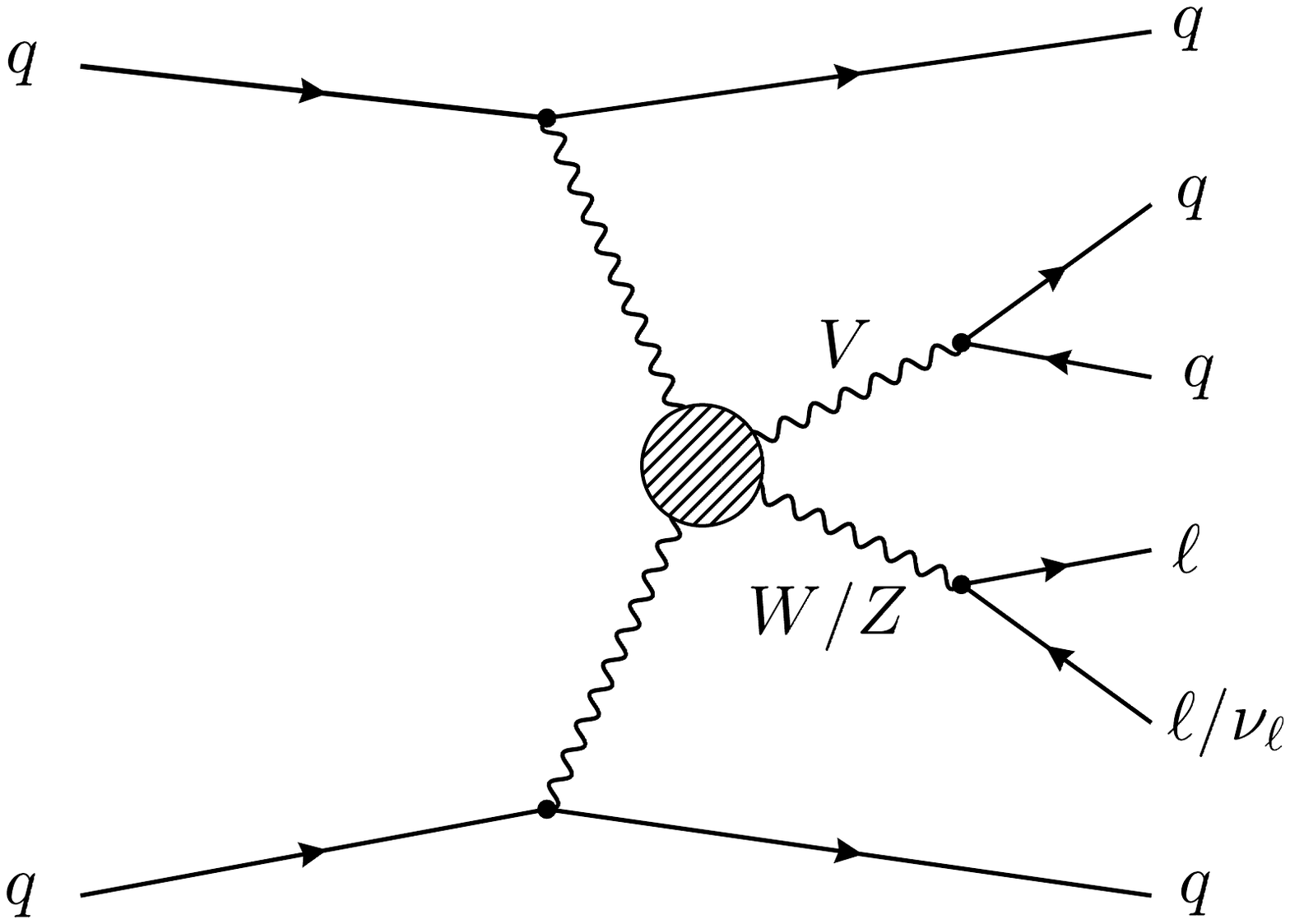}
    \includegraphics[width=0.49\textwidth]{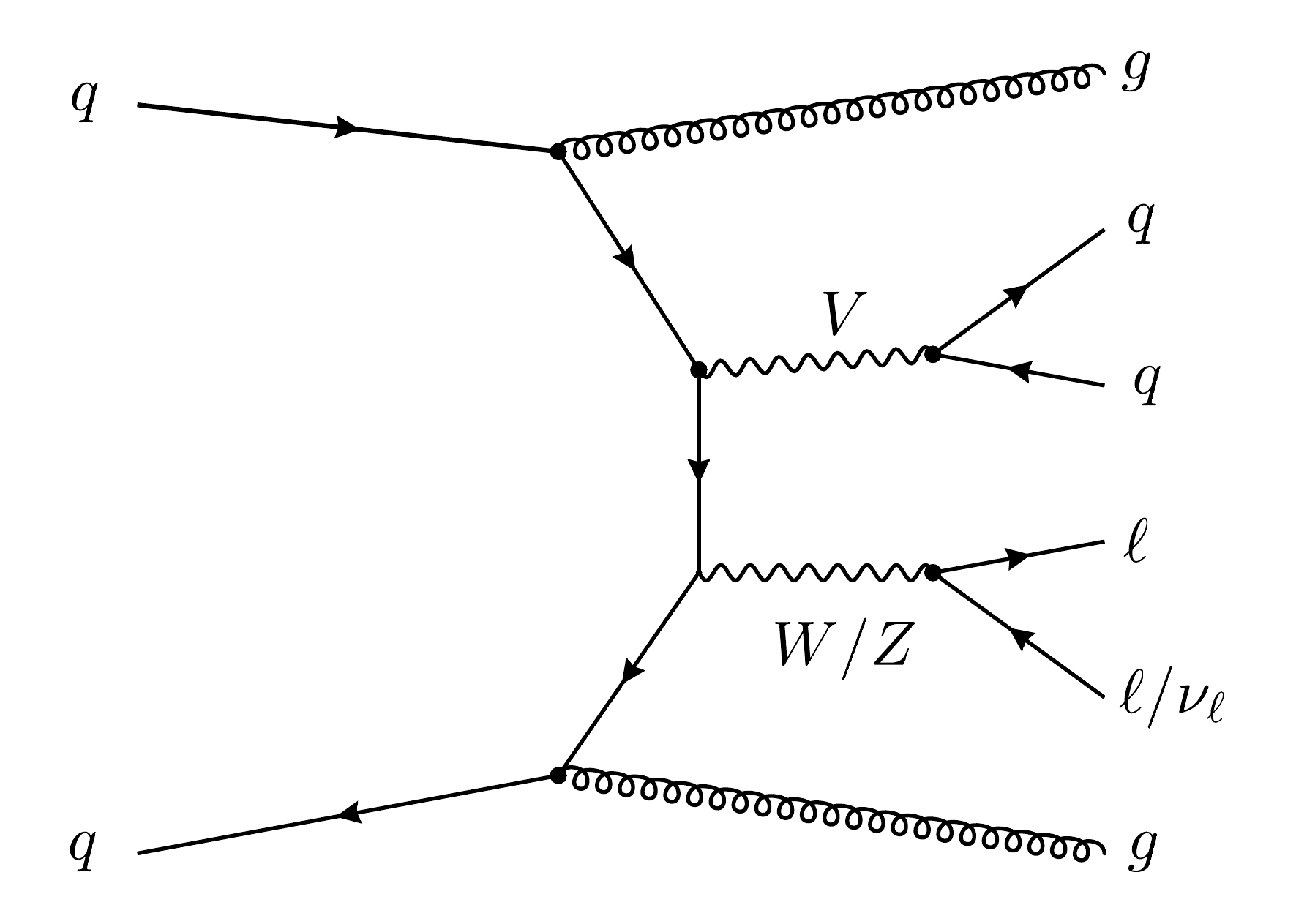}
    \caption{Examples of Feynman diagrams contributing to the analyzed final state. On the left, an illustrative diagram for the VBS process contributing to the EW production of events containing one vector boson that decays into leptons, one that decays hadronically ($\PV = \PW, \, \PZ$) and two forward jets. The BSM contributions (represented by a hatched circle) can modify the EW diboson production. On the right, an example of irreducible QCD background is presented.}
    \label{fig:vbsZVFeynman}
\end{figure}

Vector boson scattering (VBS) presents an ideal probe of the spontaneous symmetry-breaking mechanism at the CERN LHC. It provides direct access to the nonabelian gauge structure of the electroweak (EW) interactions via quartic gauge couplings and a high sensitivity to physics beyond the standard model (BSM) at energy scales higher than the direct LHC reach~\cite{Green:2016trm,Anders:2018gfr,Aaboud:2016uuk,Sirunyan:2017fvv,ATLAS:2022xyx,ATLAS:2024jpa,Covarelli:2021gyz,CMS:2021cxr,CMS:2024skm}. 
Experimentally, VBS refers to EW production of a diboson final state in association with two jets, where a constituent of each proton emits a weak boson and the two bosons scatter off each other (Figure~\ref{fig:vbsZVFeynman}). The remnants of the two protons lead to two forward/backward jets, characterized by a large rapidity separation and a large invariant mass (VBS jets).
When including the boson decays into fermions, the theoretical description of the VBS process includes, at tree level, a purely EW contribution of order $\mathcal{O}(\alpEW^6)$, where \alpEW is the EW coupling defined as $g_W^2/4\pi$, and $g_W$ is the weak coupling associated with the $\text{SU(2)}_L$ gauge symmetry of the standard model (SM). At tree level, additional irreducible contributions are present in the same final state of order $\mathcal{O}(\alpS^{2}\alpEW^4)$, where \alpS is the strong coupling, referred to as quantum chromodynamics (QCD) contributions.
 
The ATLAS and CMS Collaborations have studied VBS processes with \PW and \PZ bosons in fully leptonic final states~\cite{ATLAS:2024ett,ATLAS:2018mxa,CMS:2020etf,CMS:2020fqz,CMS:2020gfh,ATLAS:2020nlt,CMS:2022woe,ATLAS:2023sua,SSWW-CMS,SSWW-ATLAS} and also in mixed decay channels involving both leptonic and hadronic boson decays, with a subset of the LHC data taken from 2016 to 2018 (Run 2)~\cite{ATLAS:2019thr, CMS:2019qfk}. In particular, the first evidence for the $\PW\PV$ process ($\PV = \PW, \, \PZ$) in the $\Pell \PGn\PQq\PQq$ channel was presented by CMS exploiting the Run 2 data sample~\cite{CMS:2021qzz}. 
This paper, based on the full Run 2 data set, presents the first measurement of the complementary VBS process $\PZ\PV$ production, with $\PZ\to\Pell^{+}\Pell^{-}$, where \Pell is an electron or a muon, that has not yet been observed at the LHC.
This VBS signature profits from the large $\PV \to \cPq\cPaq^{(\prime)}$ branching fraction, despite having a large background contamination from other processes, such as the irreducible $\mathcal{O}(\alpS^{2}\alpEW^4)$ QCD background and the Drell--Yan (DY) processes $\mathcal{O}(\alpS^{4}\alpEW^2)$. The sensitivity to anomalous couplings is further enhanced by the use of $\PV \to \cPq\cPaq^{(\prime)}$ events where the quark jets are merged into a single large-radius jet, complementing the standard topology with two separate jets.

The data come from proton-proton collisions at $\sqrt{s} = 13\TeV$ collected from 2016 to 2018 with an integrated luminosity of 138\fbinv.

In the first part of this paper, we study the EW VBS production of $\PZ\PWpm\text{jj}$ and $\PZ\PZ\text{jj}$, where one \PZ boson decays into charged leptons ($\Pell = \Pe, \, \PGm$) and the other vector boson decays hadronically. 

In the second part of this paper we set stringent limits on BSM physics via an effective field theory (EFT)~\cite{BuchmuellerWyler, warsaw} description combining the results from this analysis with those from $\PWpm\PW\text{jj}$ -- $\PWpm\PZ\text{jj}$ production as reported in Ref.~\cite{CMS:2021qzz}	.

The theory is implemented by introducing the following effective Lagrangian:
\begin{equation}
    \mathcal{L}_{\text{eff}} = \mathcal{L}_{\text{SM}} + \sum_{D_i>4}\sum_{i}\dfrac{c_i^{(D_i)}}{\Lambda_{\text{BSM}}^{D_i -4}}\mathcal{O}^{(D_{i})}_i,
\end{equation}
where $\mathcal{L}_{\text{SM}}$ is the SM Lagrangian, the operators $\mathcal{O}_i^{(D)}$ are constructed with SM fields at some dimension $D_i$, and $c_i^{(D_{}i)}$ are the Wilson coefficients.
In this way, the contribution $\mathcal{A}_{\text{BSM}}$ to the total scattering amplitude from the EFT operators is given by:
\begin{align}
    |\mathcal{A}_{\text{BSM}}|^2 =& \sum_i^{D_i>4}\Bigg[\dfrac{c_i^{(D_i)}}{\Lambda^{D_i-4}_{\text{BSM}}}2\operatorname{Re}\Big|\mathcal{A}_{SM}^* \mathcal{A}_{\mathcal{O}_i^{(D_i)}}\Big| + \dfrac{{c_i^{(D_i)}}^2}{\Lambda^{2(D_i-4)}_{\text{BSM}}}\Big|\mathcal{A}_{\mathcal{O}_i^{(D_i)}}\Big|^2\Bigg] \nonumber \\
    &+ \sum_{j\neq k}^{D_j,D_k>4}\dfrac{c_j^{(D_j)}c_k^{(D_k)}}{\Lambda^{D_j+D_k-8}_{\text{BSM}}}\operatorname{Re}\Big|\mathcal{A}^*_{\mathcal{O}_j^{(D_j)}}\mathcal{A}_{\mathcal{O}_k^{(D_k)}}\Big|\label{eq:absm},
\end{align}
where the first summation runs over both the interference terms between the SM and one $\mathcal{O}_i^{(D)}$ operator and the quadratic contributions of the $\mathcal{O}_i^{(D)}$ operator and the second summation runs over the interference terms between two different $\mathcal{O}_i^{(D)}$ operators.

In this paper, we explore dimension-8 operators that induce anomalous quartic gauge couplings~\cite{PhysRevD.93.093013,Falkowski:2016cxu}, a focus motivated by the fact that VBS measurements typically yield weaker dimension-6 constraints than EW precision observables, Higgs, or inclusive diboson measurements. These operators lack an associated triple gauge vertex and are expected to significantly impact VBS production at large transverse momenta, where sensitivity to high-energy effects is enhanced.
A recent LHC EFT working group note~\cite{Durieux:2024zrg} distinguishes \textit{CP}-even from \textit{CP}-odd dimension-8 EFT operators. In our study, we employ the latest version of the unified \textsc{Feynrules} output (UFO) model~\cite{Almeida:2020ylr}, which uses operator definitions consistent with Ref.~\cite{Durieux:2024zrg}. We study all 20 operator structures presented in Ref.~\cite{Almeida:2020ylr}, including the T3 and T4 operators which have not previously been analyzed by the CMS Collaboration.
 
The paper is organized as follows. Section~\ref{sec:cms} describes the CMS detector, Section~\ref{sec:samples} presents the methods to produce the simulated data samples, and Section~\ref{sec:obj} describes the event reconstruction. The SM measurement for the VBS $\PZ\PV$ process is presented in Section~\ref{sec:vbs}, including the description of the selection of signal and control samples, the background estimation, and the discussion of the systematic uncertainties. The EFT interpretation of the combined lepton + jets measurements is in Section~\ref{sec:aqgc}. Section~\ref{sec:summary} contains a summary and conclusions. Tabulated results are provided in the HEPData record for this analysis~\cite{hepdata}.

\section{The CMS detector\label{sec:cms}}
The CMS apparatus~\cite{CMS:2008xjf,CMS:2023gfb} is a multipurpose, nearly hermetic detector, designed to trigger on~\cite{CMS:2020cmk,CMS:2016ngn,CMS:2024aqx} and identify electrons, muons, photons, and (charged and neutral) hadrons~\cite{CMS:2020uim,CMS:2018rym,CMS:2014pgm}. Its central feature is a superconducting solenoid of 6\unit{m} internal diameter, providing a magnetic field of 3.8\unit{T}. Within the solenoid volume are a silicon pixel and strip tracker, a lead tungstate crystal electromagnetic calorimeter (ECAL), and a brass and scintillator hadron calorimeter (HCAL), each composed of a barrel and two endcap sections. Forward calorimeters extend the pseudorapidity coverage provided by the barrel and endcap detectors. Muons are reconstructed using gas-ionization detectors embedded in the steel flux-return yoke outside the solenoid. More detailed descriptions of the CMS detector, together with a definition of the coordinate system used and the relevant kinematic variables, is reported in Refs.~\cite{CMS:2008xjf,CMS:2023gfb}.

\section{Simulated samples\label{sec:samples}}
The purely EW contributions to the $\Pell\Pell\PQq\PQq$ VBS final state are modeled using Monte Carlo (MC) simulated samples at leading order (LO) with~\MGvATNLO v2.4.2~\cite{Alwall:2014hca}, including those with $\PZ\PV$ and $\PW\PV$ intermediate states, where the first boson decays leptonically and the \PV boson decays hadronically.
The intermediate-state vector boson pair is produced by implementing the narrow-width approximation and then decayed by \textsc{MadSpin}. To preserve gauge invariance, the simulation also includes diagrams involving a \PZ boson in association with a top quark and a light quark, leading to the same $\Pell\Pell\PQq\PQq$ final states. This contribution is filtered at the event selection level and classified as $\cPqt\PZ\cPq$ background during analysis.
The QCD contribution to the $\Pell\Pell\PQq\PQq$ VBS final state is produced at LO with the \MGvATNLO~v2.4.2 generator. The interference between the EW and QCD contributions, including the terms of order $\alpS\alpEW^{5}$,
contributes less than 3\% to the inclusive cross section compared to the purely EW contributions over the phase space region of interest of the analysis~\cite{CMS:2021qzz} and is therefore neglected.
The effects of the EFT dimension-8 operators are simulated with \MGvATNLO~v2.4.2 at LO. We introduce twenty operators modifying the interaction between two scattering \PV bosons defined in Ref.~\cite{Almeida:2020ylr}. The effects of the new EFT operators are evaluated using the \MGvATNLO reweighting technique~\cite{reweight2016,Belvedere:2024nzh}. 

The background processes simulated are: (i) vector boson plus jets production (DY and {\PW}+jets); (ii) single top quark production processes in $s$-, $t$-channel, and $\cPqt\PW$; (iii) pair \ttbar and \ttbar with vector boson ($\ttbar\PV$) mechanisms; (iv) single vector boson EW production via vector boson fusion; and (v) diboson ($\PV\PGg$ and $\PV\PGg^{\star}$) and triboson ($\PW\PW\PW$, $\PW\PW\PZ$, $\PW\PZ\PZ$, $\PZ\PZ\PZ$) production. The DY and {\PW}+jets samples are generated at LO accuracy with up to four additional partons, using the MLM matching and merging scheme~\cite{Frederix:2012ps} with \MGvATNLO~v2.2.2 (v2.4.2)~\cite{Alwall:2014hca} for the 2016 (2017--2018) samples. \ttbar and single top quark processes are simulated at next-to-leading order (NLO) with \POWHEG~v2~\cite{Alioli:2010xd}. Events from \ttbar production are also weighted using generator-level information to improve the modeling of the transverse momentum distributions of the jets~\cite{Czakon:2017wor,CMS:2016oae,CMS:2019raw}.
The $\ttbar\PV$ and the vector boson fusion events are generated at LO with \MGvATNLO~v2.2.2 (v2.4.2)~\cite{Alwall:2014hca} for the 2016 (2017--2018) samples.

{\tolerance=9600
\PV boson production in association with a photon ($\PV\PGg$) is simulated with \MGvATNLO~v2.4.2 at NLO accuracy with up to 1 additional parton, using the FxFx jet merging scheme~\cite{Frederix:2012ps}. Diboson processes containing at least one \PZ boson or a virtual photon ($\PV\PGg^{*}$) with a mass as low as 100\MeV are generated with \POWHEG~v2~\cite{Nason:2013ydw} at NLO accuracy. Triboson processes with inclusive decays are also simulated at NLO accuracy with \MGvATNLO~v2.4.2.
\par}

The parton distribution function (PDF) used to generate the simulated background samples for 2016 (2017 and 2018) is the NNPDF 3.0 NLO (NNPDF 3.1 next-to-NLO)~\cite{Ball_2015,thennpdfcollaboration2017parton}.
The generators used for signal and background processes are interfaced with  \PYTHIA~8.226 (8.230) to simulate parton showering and hadronization in 2016 (2017 and 2018).
All MC samples use the \PYTHIA~\cite{Sjostrand:2014zea} program with the CUETP8M1~\cite{Khachatryan_2016} (CP5~\cite{Sirunyan_2020}) underlying event tunes for the 2016 (2017 and 2018) simulation to model parton showering, hadronization, and the underlying event.
The dipole recoil scheme is used in the parton shower simulation for the EW, QCD, and EFT dimension-8 MC contributions to the VBS final state to improve the description of the additional jet emissions~\cite{ballestrero_precise_2018,Jager:2020hkz}.

Additional collisions in the same or adjacent bunch crossings (pileup) are considered by superimposing simulated minimum bias interactions onto the hard scattering process, with a distribution matching the expectation from the luminosity profile of the collected data. Simulated events are processed with the full GEANT4-based simulation~\cite{Agostinelli:2002hh} of the CMS detector.

\section{Object and event selection\label{sec:obj}}
\label{ch:selection}

The $\Pell \PGn \PQq\PQq$ and $\Pell \Pell \PQq\PQq$ VBS analysis targets events with jets, one or two charged leptons, and a high-energy \PV decaying into a pair of quarks. The analysis distinguishes between two complementary reconstruction topologies: the resolved topology, where the \PV decay products are reconstructed as two distinct jets, and the merged topology, where high-momentum \PV bosons are clustered into a single large-radius jet. 

The CMS particle-flow (PF) algorithm~\cite{CMS:2017yfk} processes particle candidates by combining information from various subdetectors to reconstruct, identify, and measure the properties of each individual particle in the event. Events are selected by triggers for either a single lepton passing \pt thresholds of 27--35\GeV for electrons (or 24--27\GeV for muons) or a lepton pair with lower thresholds: 23\GeV (leading) and 12\GeV (subleading) for the double electron trigger, 17\GeV (leading) and 8\GeV (subleading) for the double muon trigger, and 23\GeV (leading) with 8/12\GeV (subleading) for the electron/muon trigger. Exact values depend on the data-taking period~\cite{CMS:2022uhn}.

Electrons are reconstructed by associating tracks with energy clusters in the ECAL, whereas muons are identified via tracks in the muon system. Leptons must: (i) originate from the primary vertex; (ii) meet quality selection criteria; (iii) be isolated from other event activity; and (iv) be well reconstructed using a set of criteria based on the quality of the track reconstruction, shape of calorimetric deposits, and energy flux in the vicinity of the particle's trajectory~\cite{CMS:2015xaf,CMS:2018rym}.
Multivariate discriminators are used for the lepton identification~\cite{CMS:2020uim, CMS:2018rym}. In addition, a selection based on a dedicated multivariate analysis tagger developed in Ref.~\cite{CMS:2023dsu} is added for muon candidates.

Jets are reconstructed using the anti-\kt algorithm~\cite{Cacciari:2008gp, Cacciari:2011ma} with a distance parameter of 0.4 (AK4) and 0.8 (AK8). AK4 and AK8 jets are required to satisfy $\pt > 30 \GeV$ and $\abs{\eta} < 4.7$, or $\pt>200\GeV$ and $\abs{\eta}<2.4$, respectively. Reconstructed jets cannot overlap with isolated leptons: $\Delta R(\text{j}, \Pell)= \sqrt{\smash[b]{(\Delta\eta)^2 + (\Delta\phi)^2}} >$ 0.4 (0.8) for AK4 (AK8) jets. In addition, AK4 jets overlapping with AK8 jets with $\Delta R(\text{j}_{\text{AK4}}, \text{j}_{\text{AK8}}) < 0.8$ are removed from the event. Jet energy calibrations account for neutral pileup contributions and nonlinear detector responses in both simulation and data. Jet energy corrections are propagated to the missing transverse momentum (\ptmiss). The pileup contribution to AK4 jets is mitigated using charged-hadron subtraction~\cite{Cacciari:2007fd}, removing the energy of charged hadrons not originating from the primary vertex. AK8 jets are reconstructed using the \textsc{puppi} algorithm~\cite{Bertolini:2014bba} for pileup mitigation. 

The boson candidate's mass ($m_{\PV}$) is computed after applying the modified mass-drop tagger, known as ``soft drop'' (SD) algorithm~\cite{Dasgupta:2013ihk, Larkoski:2014wba}, and its substructure is quantified using $N$-subjettiness~\cite{Thaler:2010tr}. The SD algorithm, which has as an angular exponent $\beta = 0$, soft cutoff threshold $z_{\text{cut}} < 0.1$, and characteristic radius $R_{0} = 0.8$~\cite{Larkoski:2014wba, CMS:2019kuk}, removes soft, wide-angle radiation from the large radius jet, improving the modeling of the jet mass observable. The ratio of the 2- to 1-subjettiness~\cite{Thaler:2010tr}, $\tau_2 / \tau_1$, is used to distinguish hadronic boson decays from jets originating from light quarks or gluons. 
To capture the decay products of the \PV boson, events are categorized into two mutually exclusive topologies: resolved and merged. In the resolved category, no AK8 jets must be present, and at least four AK4 jets are required. For resolved events with multiple \PV candidates, the pair of AK4 jets with an invariant mass closest to the \PW or \PZ boson mass is selected. In the merged category, events contain at least one AK8 jet and two AK4 jets. This configuration corresponds to the case where the \PV decay products are highly collimated and reconstructed as a single jet.

The VBS jets, originating from initial-state radiation (ISR), are identified by selecting the pair of AK4 jets with the highest dijet mass in the event. The dijet system must satisfy $m_{\text{jj}} > 500\GeV$ and $\abs{\Delta \eta_{\text{jj}}} > 2.5$.
The deep neural network (DNN)~\textsc{DeepCSV} tagger~\cite{CMS:2017wtu, CMS:2018lgn} identifies AK4 jets likely originating from bottom quarks in the region $\abs{\eta}<2.4$. This algorithm exploits tracks, neutral particles, and secondary vertex information. The analysis applies a tagger score with approximately 85\% \PQb jet identification efficiency and a mistag rate of less than 10\% for light quark and gluon jets. 

The $\PZ\PV$ selection targets events containing two same-flavor, opposite-charge leptons, and multiple jets. 
Lepton candidates are required to be isolated, same-flavor (different-flavor in the case of the top quark enriched control region, CR), and of opposite charge, with $\pt > 35 \, (20) \GeV$ for the leading (subleading) lepton, exploiting the harder kinematic spectrum of the signal with respect to the main backgrounds, and $\abs{\eta} < 2.5 \, (2.4)$ for electrons (muons). Additionally, the invariant mass of the dilepton system must lie between 76\GeV and 106\GeV to ensure that the leptons originate from a \PZ boson decay~\cite{ParticleDataGroup:2024cfk}.
Following these preselection criteria, signal regions (SRs) are defined, as well as CRs for the main backgrounds, \eg, top quark and DY processes, within both the resolved and merged categories. The reconstructed mass of the centrally produced \PV boson is a key discriminant variable used to define these regions. In the resolved topology, this is evaluated as the invariant mass of the two AK4 jets from \PV decay, whereas in the merged topology, it is the SD mass of the leading AK8 jet. To define the SR, events are selected with $65 < m_{\PV} < 105\GeV$, corresponding to the on-shell mass window of the \PV boson. For the DY CR, this selection is inverted, since the background dominates in the off-shell region. This selection yields a DY CR purity of up to 86\%. The SR and DY CR are further subdivided based on the presence of a \PQb-tagged jet, resulting in ``\PQb veto'' and ``\PQb tag'' subcategories. Top quark CRs are defined by selecting events where the flavors of the two leptons do not match, enhancing the contribution from top quark background processes and resulting in a ${\approx} 90\%$ pure top quark sample. Table~\ref{tab:zv selection} summarizes the definition of various regions of the $\PZ\PV$ analysis.
\begin{table}[h!t]
\centering
\topcaption{Definitions of the SRs and the CRs for the VBS $\PZ \PV$ analysis for both resolved and merged categories. Lepton candidates are required to be of opposite charge.}\label{tab:zv selection}
\resizebox{\textwidth}{!}{
\begin{tabular}{cccc}
Preselection & \multicolumn{3}{c}{
\begin{tabular}[c]{@{}c@{}}
$n_{\Pell} = 2$, $m_{\Pell \Pell} \in [76, 106]\GeV$, $\pt (\Pell_{1}) > 35\GeV$, $\pt (\Pell_{2}) > 20\GeV$,\\
	$\pt (\text{j}_{1,2}) > 30\GeV$, $m_{\text{jj}} > 500\GeV$, $\abs{\Delta \eta_{\text{jj}}} > 2.5$
\end{tabular}} \\
[\cmsTabSkip]
Regions & Variables & Lepton Selection & Other Requirements \\
\hline
SR & $65 < m_{\PV} < 105\GeV$ & Same-flavor & Split in \PQb tag and \PQb veto \\
DY CR & $m_{\PV} < 65$ or $m_{\PV} > 105\GeV$ & Same-flavor & Split in \PQb tag and \PQb veto \\
Top quark CR & $65 < m_{\PV} < 105\GeV$ & Different-flavor & \NA \\
\end{tabular}
}
\end{table}

The $\PW\PV$ selection reconstructs $\Pell\PGn\PQq\PQq$ events from $\PWpm\PZ\text{jj}$ -- $\PW\PW\text{jj}$ production, and is employed in the EFT studies presented in Section~\ref{sec:aqgc}. Candidate events are required to contain exactly one charged and isolated lepton, \ptmiss higher than 30\GeV, and multiple jets. The missing transverse momentum vector, \ptvecmiss, corresponds to the negative vector sum of the transverse momenta of all particle candidates in the event. Events are divided into muon or electron categories based on the flavor of the charged lepton. Only the merged categories are considered for this final state as they are the most sensitive to dimension-8 EFT operators~\cite{Covarelli:2021gyz}. Additional details on the $\PW\PV$ selection are described in detail in Ref.~\cite{CMS:2021qzz} and summarized in Table~\ref{tab:wv selection}.

\begin{table}[h!t]
\centering
\topcaption{Definitions of the SRs and the CRs for the VBS $\PW \PV$ analysis for the merged categories. The transverse mass of
	the leptonically decaying \PW, defined as $\mT^{\PW}= \sqrt{\smash[b]{2 \pt(\Pell) \ptmiss[1-\cos (\Delta \varphi[\vec{\Pell}, \ptvecmiss])]}}$, where $\pt(\Pell)$ is the \pt of the lepton and $\Delta \varphi(\vec{\Pell}, \ptvecmiss)$ is the azimuthal distance between the lepton and the \ptvecmiss.}\label{tab:wv selection}
\resizebox{0.9\textwidth}{!}{
\begin{tabular}{ccc}
Preselection & \multicolumn{2}{c}{
\begin{tabular}[c]{@{}c@{}}
$n_{\Pell} = 1$, $\pt (\Pe) > 30\GeV$  (2016), 35\GeV  (2017, 2018),\\ $\pt(\PGm) > 30\GeV$, $\ptmiss > 30\GeV$, $\mT^\PW < 185\GeV$, \\ 
$\pt(\text{j}_{1}) > 50\GeV$, $\pt(\text{j}_{2}) > 30\GeV$, $m_{\text{jj}} > 500\GeV$, $\Delta \eta_{\text{jj}} > 2.5$
\end{tabular}} \\
[\cmsTabSkip]
Regions & Variables & Other Requirements \\
\hline
SR & $70 < m_{\PV} < 115\GeV$ & \PQb veto \\
{\PW}+jets CR & $m_{\PV} < 70\GeV$ or $m_{\PV} > 115\GeV$ & \PQb veto \\
Top quark CR & $70 < m_{\PV} < 115\GeV$ & \PQb tag \\
\end{tabular}
}
\end{table}

\section{Measurements of the Standard Model VBS \texorpdfstring{\PZ{}\PV}{ZV} process\label{sec:vbs}}
\subsection{Background estimation and analysis strategy}
\label{sec:strategy}
The primary background sources for the $\PZ\PV$ VBS signature are DY and top quark production processes. The DY background occurs when two leptons are produced alongside jets from ISR, whereas top quark backgrounds arise from processes such as top quark pair or single top quark production.

\begin{figure}
\centering
\includegraphics[width=0.49\textwidth]{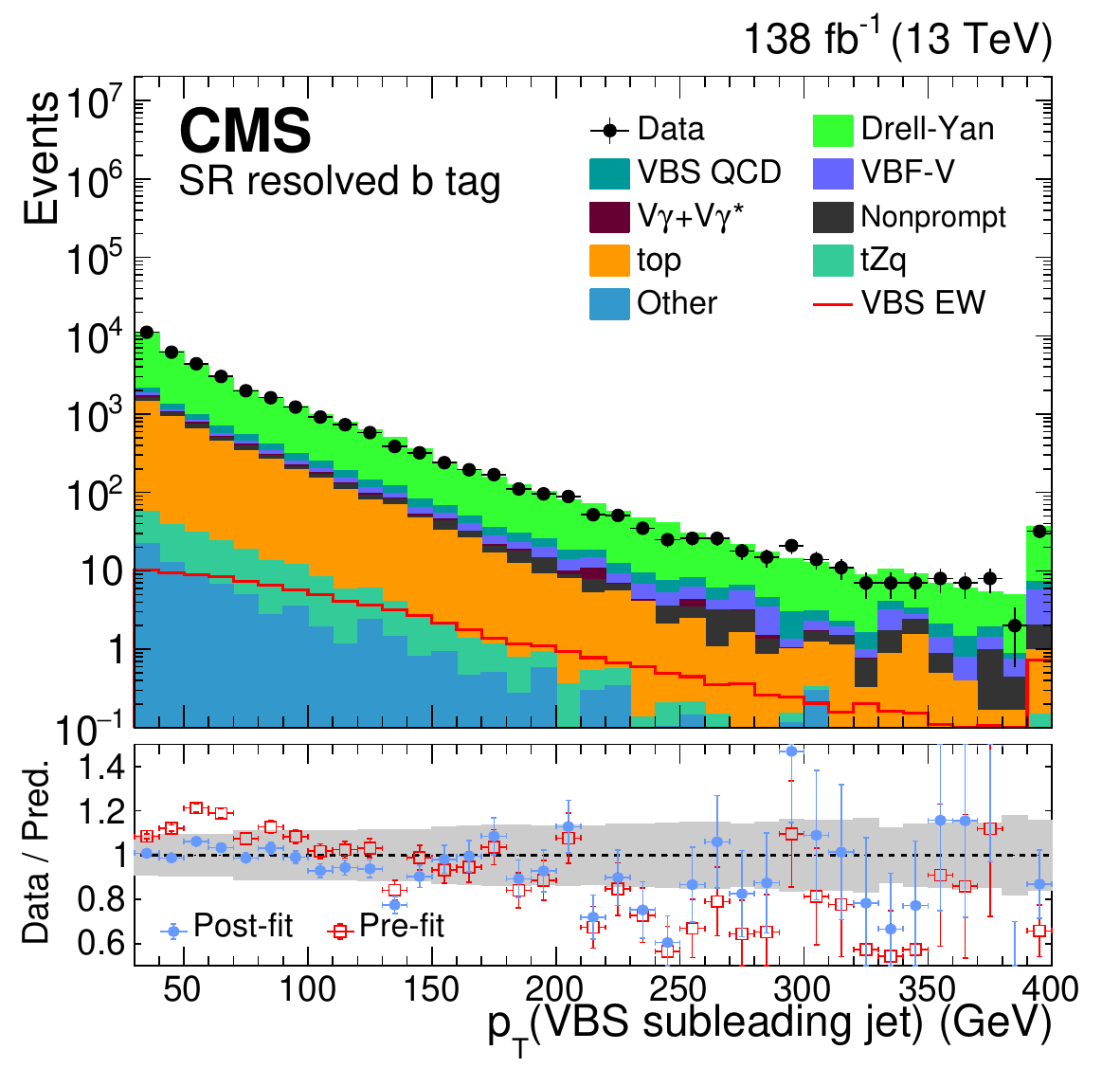}
\includegraphics[width=0.49\textwidth]{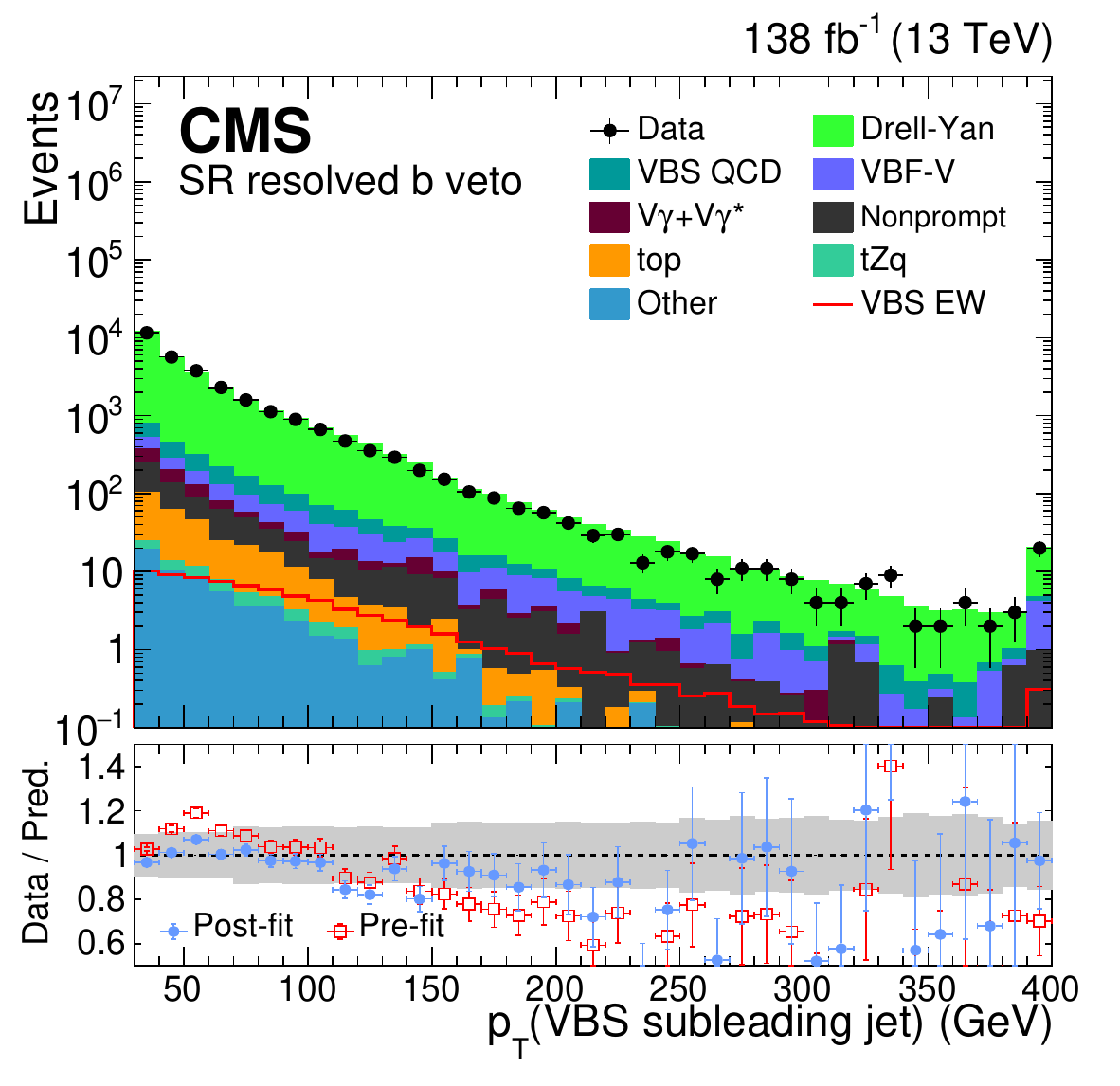}
\caption{Distributions of subleading VBS jet \pt for the data and post-fit backgrounds (stacked histograms), in the SRs of the $\PZ \PV$ channel for the resolved \PQb tag (left) and the resolved \PQb veto (right) categories. The post-fit VBS EW $\PZ\PV$ signal is shown overlaid as a red solid line. The overflow is included in the last bin. The lower panels show the ratios of the data to the pre-fit background prediction and post-fit background yield as red open squares and blue points, respectively. The gray band in the lower panels indicates the systematic component of the post-fit background uncertainty. The vertical bars on the data points represent statistical uncertainties.}\label{fig:paper SR jet pt2}
\end{figure}

The estimation of the DY contribution relies on MC simulation, corrected using observed events. This method addresses both the mismodeling of jet momentum distributions~\cite{Lindert:2017olm,CMS:2019raw,CMS:2021qzz} and missing higher-order effects in QCD and EW perturbative expansions. The DY samples are categorized based on kinematic variables that exhibit poor modeling, and a rate parameter is assigned to each bin of the DY MC prediction. The normalizations of these contributions are allowed to vary freely in the simultaneous maximum likelihood (ML) fit. For the resolved categories in 2017 and 2018, where mismodeling is more pronounced due to a combination of detector conditions and the limited statistical precision of the 2016 samples, DY samples are divided into 12 sub-regions defined by the \pt of the leptonically decaying \PZ boson and subleading VBS jet. In contrast, the resolved category for 2016 and merged categories are divided into 5 bins, based only on the $\pt$ of the leptonically decaying \PZ boson. Figure~\ref{fig:paper SR jet pt2} illustrates the \pt distributions of the subleading VBS jet, a variable notably affected by modeling discrepancies, comparing data with pre-fit and post-fit background estimates derived from the simultaneous ML fit of the SRs and CRs.

The top background shape comes from simulation, but its normalization is measured from data collected in the dedicated CR. The normalizations of the top quark backgrounds are extracted from the simultaneous fit across the SRs and CRs.
Other sources of SM backgrounds arising from {\PW}+jets and multiple bosons production account for a smaller contribution in the signal phase space. They come directly from MC simulations, with normalization derived from their theoretical cross sections. Finally, the QCD multijet background, which may enter the SR with nonprompt leptons, is estimated from data by measuring the probability for a lepton originating from a jet to be misidentified as an isolated lepton in a phase space region outside the analysis region. This auxiliary region is defined from events in which one lepton satisfies the standard identification and isolation criteria, and a second lepton candidate fails these criteria but passes a relaxed set of requirements. A detailed description of the nonprompt background estimation is given in Refs.~\cite{CMS:2022uhn,CMS:2018zzl}.

{\tolerance=1200
Because of the large background and complex signal topology, the most significant kinematic variables to separate signals and backgrounds are used as input to a single discriminator built with a fully connected deep neural network (DNN). 
The DNN is implemented with the \textsc{Keras}~\cite{keras} framework, which provides a high-level interface for model development, and uses \textsc{TensorFlow}~\cite{tensorflow2015-whitepaper} to perform the underlying numerical computations required for training and evaluation. It is trained specifically to enhance the separation of the VBS EW signal from all background processes.
\par}

Separate DNN models are trained for each topology (resolved or merged) and for each \PQb tagging region (\PQb-tagged or \PQb-vetoed) to discriminate the signal against all backgrounds at the same time, providing a continuous output, allowing a binary classification to be performed.

Because of the samples available for training, a different training for each data-taking year is performed in the resolved category, whereas in the merged category, inclusive training across all three years was chosen due to the smaller data set. This is achieved by combining all simulation samples into a single dataset containing the information on the cross section, selection, and scale factor coefficients applied as a weight on a per-event basis. The training is performed in the SR defined in the previous section, with 80\% of the events used for training and 20\% reserved for testing.
The architecture selected is a dense neural network that uses binary cross-entropy as the loss function and the \textsc{Adam} optimizer~\cite{kingma2017adam} for gradient descent. 

All hidden layers employ the Rectified Linear Unit~\cite{agarap2018deep} activation function and both L1 and L2 kernel regularization~\cite{Goodfellow-et-al-2016} are applied. Dropout is also added to mitigate overfitting, together with an early stopping criterion. The output layer is a single neuron activated by a sigmoid function. The hyperparameters (\eg, learning rate, regularization parameters, dropout fraction) were optimized using Bayesian optimization~\cite{bayes-opt}.
To limit the impact of potentially poorly modeled input variables, the feature set is pruned iteratively based on the ranking of their importance measured with an estimator from the SHAP (SHapley Additive exPlanations)~\cite{shap} technique. Specifically, the two least important features are removed at each iteration, and the entire training and optimization procedure is repeated. The final pruned model is selected as the smallest set of variables that does not exhibit a significant loss of performance.

Table~\ref{tab: DNN rankings} shows the list of input variables used for the four categories of the analysis.
The sets of input variables for the DNNs include quantities suited for the particular kinematical properties of the EW VBS processes. Among these are the event Zeppenfeld $\zeta$ variable~\cite{Rainwater:1996ud} of the leptons or merged jet in the EW VBS processes with respect to the scattered VBS jets, and the quark/gluon discriminator variable, or quark-gluon likelihood (QGL), of the leading \PV jet, which is based on a likelihood discriminant constructed with three variables for each jet: the jet energy, the multiplicity of the jet constituents, and the minor axis width of the ellipse in the $\eta$ - $\phi$ plane containing the jet constituents~\cite{CMS:2013kfa,CMS-DP-2016-070}. In particular, the most important input variables for each category consistently include the dijet invariant mass as the highest-ranking feature, followed by either the number of jets, the dijet $\eta$ separation, or the leading lepton $\zeta$, depending on the specific topology and tagging region.

\begin{table*}[htbp]
    \centering
    \topcaption{Variables used as input to the DNN for the resolved and merged models in the two \PQb-tagged categories. The check mark indicates that the variable is included in the DNN model identified in the column header.}
    \label{tab: DNN rankings}
      \begin{tabular}{l l l l l}
      \multirow{3}{*}{Feature} & \multicolumn{4}{c}{Category}\\\cline{2-5}
                                 & \multicolumn{2}{c}{Resolved} & \multicolumn{2}{c}{Merged}\\\cline{2-5}
                                 & \PQb veto & \PQb tag & \PQb veto & \PQb tag\\
        \hline
        Leading lepton \pt & \NA & \checkmark & \checkmark & \checkmark\\
        Leading lepton $\eta$ & \NA & \NA &  \checkmark & \NA \\
        Subleading lepton $\eta$ & \NA & \NA & \checkmark & \NA \\
        Dilepton mass & \NA & \checkmark & \NA & \NA \\
        Leading lepton $\zeta$ & \checkmark & \checkmark  &  \checkmark & \checkmark \\
        Subleading lepton $\zeta$ & \checkmark & \checkmark & \checkmark & \checkmark \\
        Subleading VBS jet \pt & \checkmark & \checkmark  & \checkmark & \checkmark  \\
        Leading VBS jet $\eta$ & \NA & \NA & \NA & \checkmark \\
        Subleading VBS jet $\eta$ & \NA & \NA & \checkmark  & \checkmark  \\
        Dijet mass & \checkmark  & \checkmark  & \checkmark  & \checkmark  \\
        Dijet $\eta$ separation & \checkmark  & \checkmark  & \checkmark  & \checkmark \\
        Dijet $\phi$ separation & \checkmark  & \checkmark  & \checkmark  & \NA\\
        Number of jets ${>}30\GeV$ & \checkmark  & \checkmark  & \checkmark  & \checkmark  \\
        Number of \PQb-tagged jets & \NA & \NA & \NA & \checkmark \\
        Leading \PV jet \pt & \checkmark  & \checkmark   & \NA & \NA \\
        Subleading \PV jet \pt & \checkmark  &  \checkmark & \NA & \NA \\
        Leading \PV jet $\eta$ & \checkmark & \checkmark & \NA & \NA \\
        Subleading \PV jet $\eta$ & \checkmark  & \checkmark   & \NA & \NA \\
        Leading \PV jet QGL & \checkmark  & \checkmark  & \NA & \NA \\
        Subleading \PV jet QGL & \checkmark  & \NA & \NA & \NA \\
        Merged jet \pt & \NA & \NA & \checkmark  & \checkmark  \\
        Merged jet $\eta$ & \NA & \NA & \checkmark  & \checkmark  \\
        Merged jet $\zeta$ & \NA & \NA & \checkmark  & \checkmark \\
        \PV jet mass & \checkmark  & \checkmark & \NA & \NA \\       
      \end{tabular}
\end{table*}

The measurement of the SM VBS processes under investigation is performed using a statistical analysis implemented via a ML fit. The signal strength, defined as the ratio of the observed signal yield to that predicted by the model, is extracted, with its uncertainty determined under the asymptotic limit~\cite{wilks} via Wilks' theorem. The ML fit is performed using the CMS statistical analysis tool \textsc{Combine}~\cite{CMS:2024onh}, which is built on the \textsc{RooFit}~\cite{Verkerke:2003ir} and \textsc{RooStats}~\cite{Moneta:2010pm} frameworks. Data yields from both the SRs and CRs are incorporated into the likelihood through Poisson probability density functions. The SR inputs to the fit include the DNN output distributions for the data, signal, and background, estimated as previously described.

\subsection{Systematic uncertainties}\label{sec:zvsystematics}
Sources of systematic uncertainty affecting the DNN distribution and yield normalization include statistical uncertainties in experiment and simulation, together with systematic uncertainties in quantities affecting the modeling in simulation. The latter are treated as free parameters in the fitting procedure, constrained by log-normal probability density functions inserted into the likelihood function to characterize their uncertainties. 

The uncertainties in the CMS integrated luminosity measurement (1.2\%~\cite{lumi16}, 2.3\%~\cite{lumi17}, and 2.5\%~\cite{lumi18} for 2016, 2017, and 2018, respectively) are partially correlated between the data-taking years, resulting in an overall uncertainty of 1.6\%. These uncertainties affect only the integrated yields, without effects on the shapes of the distributions. Systematic uncertainties related to the pileup modeling are introduced as a ${\pm} 4.6\%$ variation of the total inelastic cross section of 69.2 mb that is used to estimate the data pileup distributions~\cite{CMS:2018mlc}. They are considered uncorrelated among the data-taking years. The lepton trigger, reconstruction, identification, and isolation efficiencies are measured in both experiment and simulation using $\PZ \to \Pell \Pell$ events. Data-to-simulation scale factors are applied to all simulation samples to account for differences. The uncertainties in the scale factors are included by changing their values by ${\pm}1$ standard deviation from the nominal values and are treated as uncorrelated among data-taking years. 

In 2016 and 2017, a portion of trigger primitives in the ECAL was associated with the wrong bunch crossing, leading to a trigger mistiming and a nonnegligible decrease in the trigger efficiency that is not modeled in the simulated samples~\cite{CMS:2024ppo}. Events have been corrected for the mistiming efficiency loss with a per-event weight, and the corresponding uncertainties have been propagated throughout the analysis chain. The resulting uncertainty in the signal yields amounts to less than 2\%. They are considered uncorrelated among the data-taking years. Jet energy scale and resolution uncertainties are evaluated by shifting the \pt value of the jets, thus directly affecting the reconstructed jet multiplicity measurement~\cite{Khachatryan:2016kdb}; several independent sources are considered and partially correlated among different data sets. These uncertainties are included for both AK4 and AK8 jets. 

For the merged category, \PV tagging corrections with their corresponding uncertainties are included. These variables are calibrated in a top quark-antiquark sample enriched in hadronically decaying \PW bosons~\cite{CMS:2014rsx}. Uncertainties in \PQb tagging and mistagging data-to-simulation scale factors are applied to reproduce the corresponding efficiencies measured in the data, and implemented in the ML fit as correlated among the data-taking years.

The dominant theoretical uncertainty arises from the choice of renormalization and factorization scales in the MC event simulation. This uncertainty is evaluated by taking the largest variation obtained by independently changing the nominal value of each scale up and down by a factor of two, excluding the two most extreme variations~\cite{Kalogeropoulos:2018cke}. For DY and top quark backgrounds only shape effects are included since their normalization is directly measured from observed events in the fit. Both the shape and normalization effects are included for other backgrounds. This uncertainty is considered uncorrelated among the different classes of processes and correlated among the data-taking years. The uncertainty in the modeling of the parton shower is also included, by using the weights corresponding to variations of the strong coupling \alpS in the description of ISR and final-state radiation in the parton showering programs, and is treated as uncorrelated among various processes. The PDF and related strong coupling \alpS uncertainties are evaluated from the eigenvalues of the PDF set following the NNPDF prescription~\cite{Rojo:2016ymp}. These uncertainties, as well as the ones from the modeling of the underlying event, are included for all processes apart from the top quark and DY backgrounds, and they have a negligible impact on the signal measurement. Finally, a systematic uncertainty is applied to the top-quark \pt reweighting~\cite{Czakon:2017wor,CMS:2016oae,CMS:2019raw}.

For the signal and background processes estimated from simulation, the limited MC sample size limits the precision of the modeling. The corresponding statistical uncertainties are estimated using the Barlow--Beeston lite method~\cite{BB}, and are therefore taken as systematic uncertainties applied to each bin of the corresponding distribution.

The estimate of the nonprompt background is affected by an irreducible uncertainty due to the limited number of events in the auxiliary regions used to extrapolate it in the main regions. This uncertainty is treated similarly to the statistical uncertainty assigned to the other estimates obtained from simulations, introduced above. Furthermore, we assign to the nonprompt background also a 30\% uncertainty in the integrated yields to account for all possible uncertainties related to its data-driven estimate. The evaluation of this uncertainty is derived from a closure test to its estimate from CRs in data, where this value resulted to be the maximum relative discrepancy between the nonprompt estimate and observation. It is treated as correlated among the data-taking years.

The impacts of the systematic uncertainties on the EW $\PZ\PV$ VBS signal strength, as extracted from the ML fit, are summarized in Table~\ref{tab:systUncewZVrun2}. We have checked the sensitivity of the results to variations of the individual nuisance parameters; no significant overconstraining or underestimating of the systematic uncertainties was found.

\begin{table}[htb]
\centering
\caption{The impact of each systematic uncertainty, together with the impact of the collected data statistical uncertainty, on the signal strength $\mu$, as extracted from the fit to measure the EW $\PZ\PV$ VBS signal with the DNN output distributions. Upper and lower uncertainties are given for the various sources. The theory uncertainty includes contributions from both the signal and backgrounds.}
\begin{tabular}{l>{$}c<{$}>{$}c<{$}}
Source & - \Delta \mu & + \Delta \mu\\
\hline
Experimental                                        & -0.27 & +0.28 \\
\quad DY and top quark bkg. normalizations        & -0.20 & +0.20  \\
\quad MC sample size (bin-by-bin unc.)   & -0.14 & +0.15  \\
\quad Other nuisances                   & -0.11 & +0.13   \\
Theory                            & -0.24 & +0.26  \\
Statistical                   & -0.36 & +0.36  \\
[\cmsTabSkip]
Total                             & -0.51 & +0.53  \\
\end{tabular}
  \label{tab:systUncewZVrun2}
\end{table}

\subsection{Results}

\begin{figure}[htpb!]
    \centering
    \includegraphics[width=0.49\linewidth]{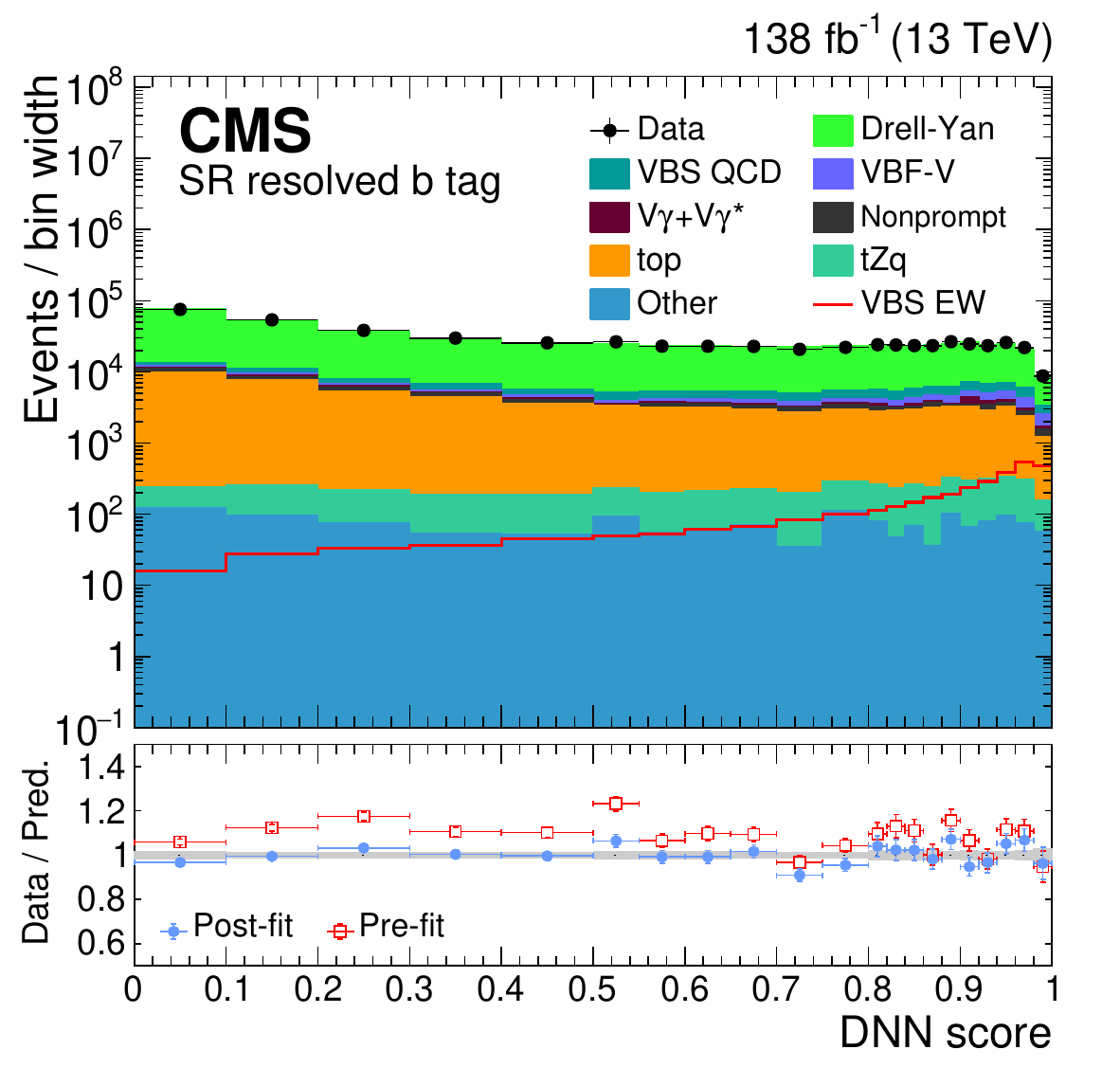}
    \includegraphics[width=0.49\linewidth]{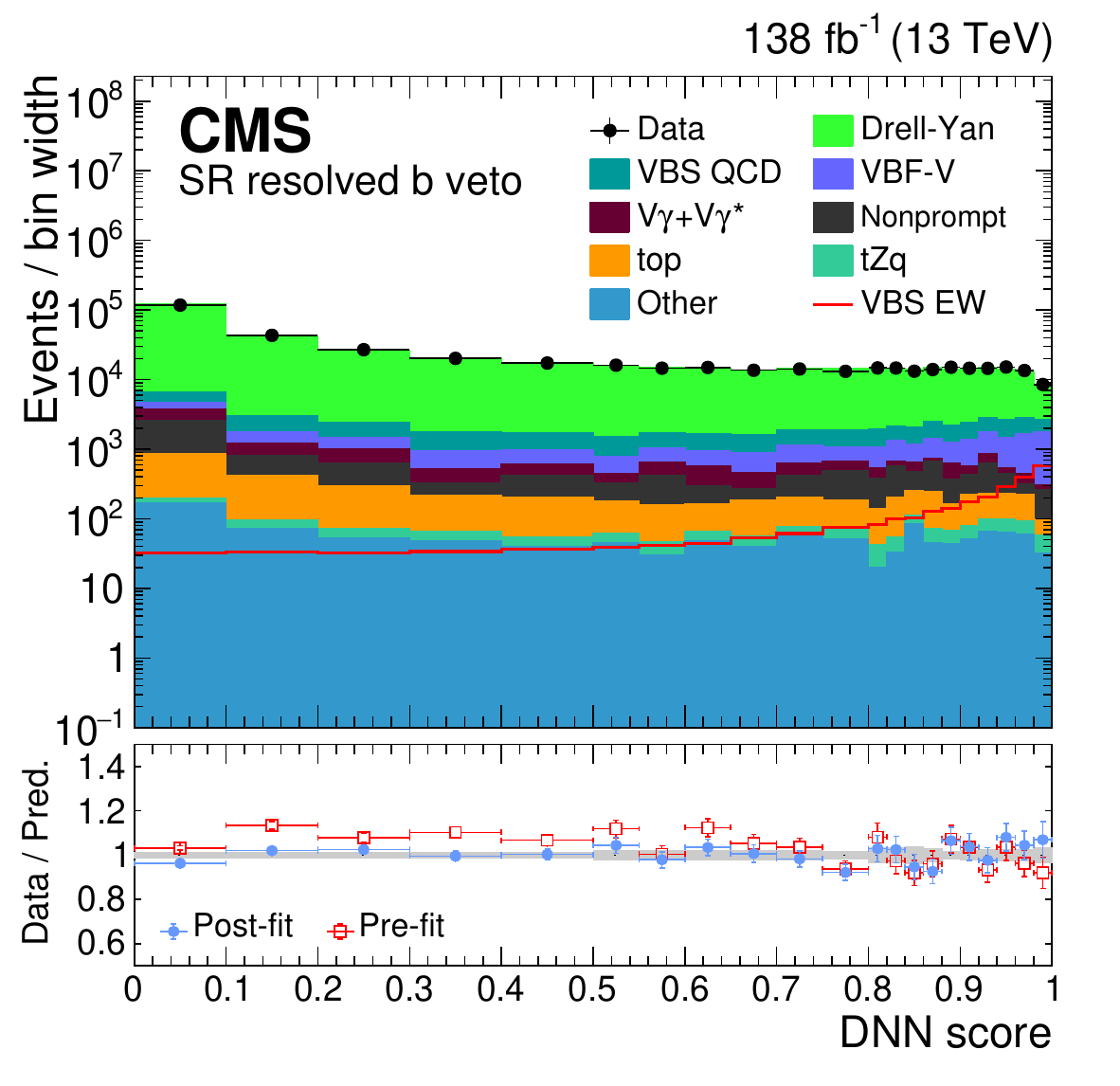}\\ 
    \includegraphics[width=0.49\linewidth]{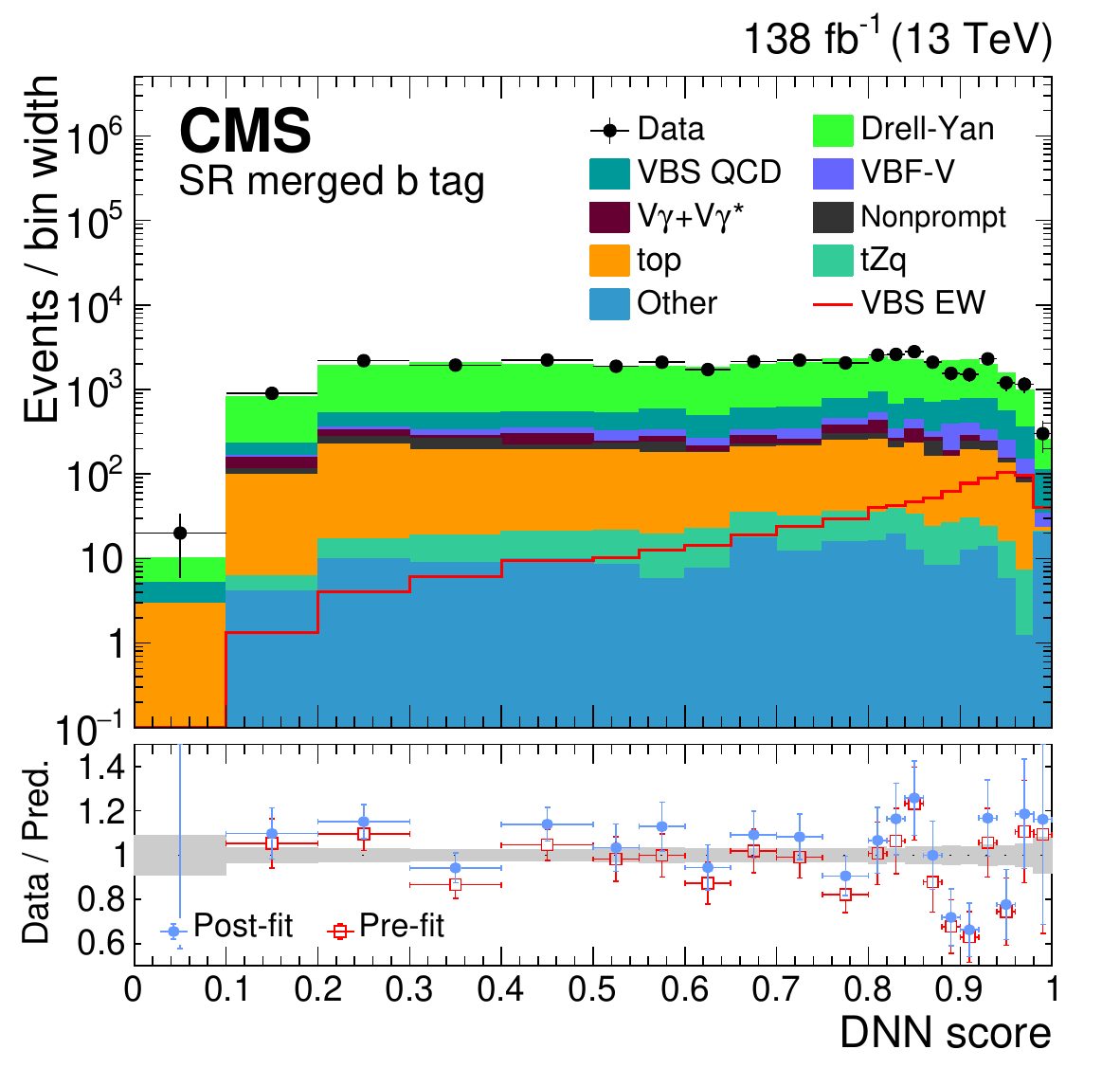}
    \includegraphics[width=0.49\linewidth]{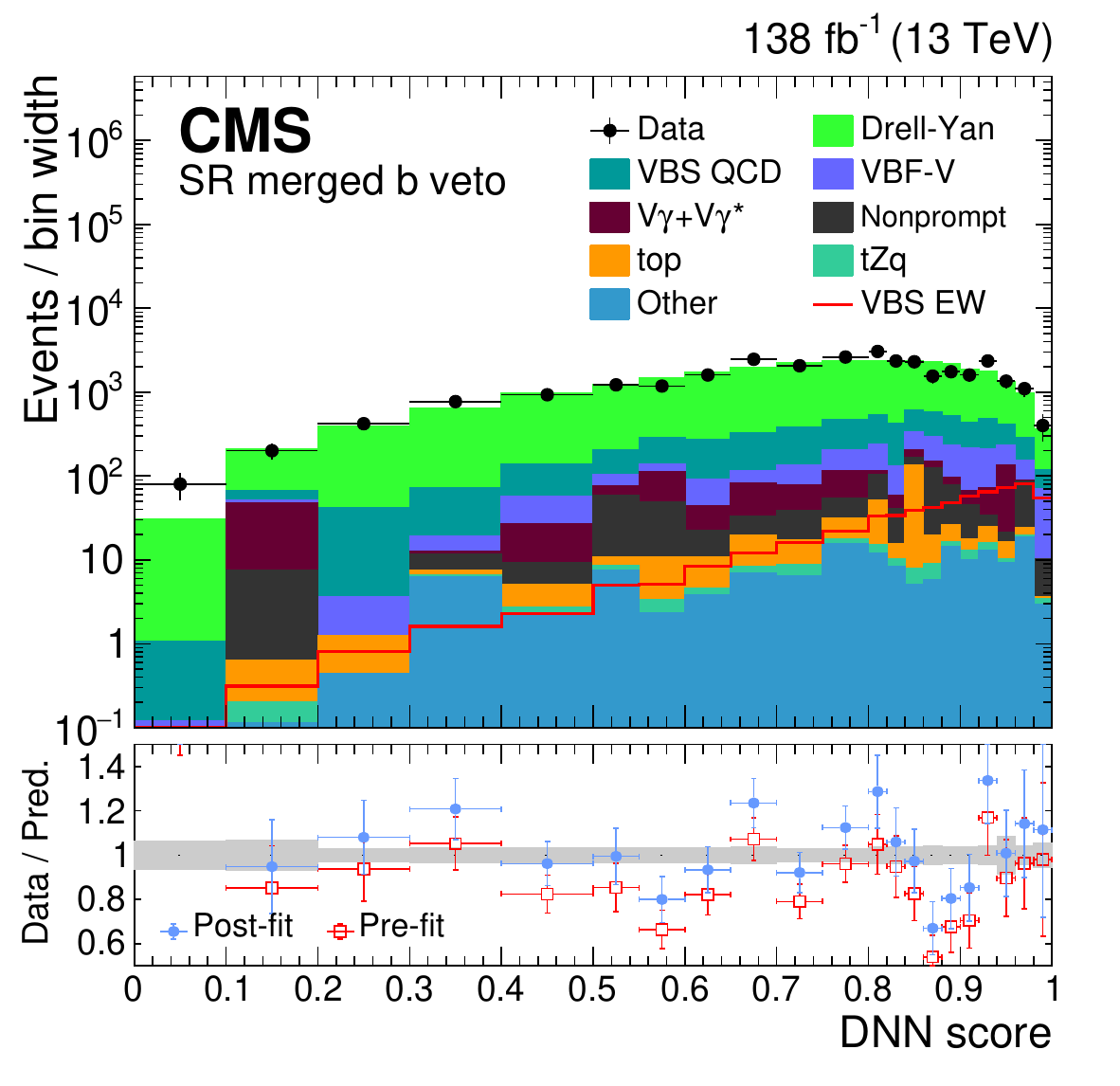}
    \caption{Distributions of DNN score for the data and post-fit backgrounds (stacked histograms), in the SRs of the $\PZ\PV$ channel for the \PQb tag (left) and the \PQb veto (right) channels, for the resolved (merged) category in the first (second) row. The post-fit VBS EW $\PZ\PV$ signal is shown overlaid as a red solid line. The overflow is included in the last bin. The lower panels show the ratios of the data to the pre-fit background prediction and post-fit background yield as red open squares and blue points, respectively. The gray band in the lower panels indicates the systematic component of the post-fit background uncertainty. The vertical bars on the data points represent statistical uncertainties.}
 \label{fig:postfit_DNN_SM_RunII}
\end{figure}

We extract values of the signal strength and the statistical significance from the fit of the $\PZ\PV$ signal.  We consider the purely EW signal strength keeping the QCD $\PZ\PV$ production contribution fixed to the SM prediction. For the primary result of the EW signal strength measurement, the DNN output distributions of the SRs and the top quark and DY CRs are shown in Figure~\ref{fig:postfit_DNN_SM_RunII} for both the resolved and merged topologies. The data are compared with the background estimated before (pre-fit) and after (post-fit) the simultaneous fit of the SRs and CRs. 

The observed (expected) EW signal strength is $0.63^{+0.53}_{-0.51}$ ($1.00^{+0.61}_{-0.58}$), including theory uncertainties from both the signal and backgrounds, corresponding to a signal significance of 1.3 (1.8) standard deviations with respect to the null hypothesis.
The largest contribution to the overall uncertainty is the statistical uncertainty in the data, as reported in Table~\ref{tab:systUncewZVrun2}.

\section{The EFT interpretation of \texorpdfstring{$\Pell\PGn\PQq\PQq$}{ellnuqq} and \texorpdfstring{$\Pell\Pell\PQq\PQq$}{ellellqq} VBS data}
\label{sec:aqgc}

The investigation of EFT contributions to the VBS processes of interest also relies on a ML fit, and combines the data from the $\PZ\PV$ and $\PW\PV$ channels.
The expected number of events, $N_{\textrm{exp}}$, inherits a quadratic dependence on the EFT Wilson coefficients from the square of the scattering amplitude $\abs{\mathcal{A}_{\text{BSM}}}^2$ as described in Eq.~\eqref{eq:absm}. When considering a single EFT dimension-8 operator (associated with the Wilson coefficient $f_\alpha$), with all others set to zero, the expected number of events is:
\begin{equation}\label{eq:1D8}
N_{\textrm{exp}}^{\alpha} = N_\mathrm{SM} + \frac{f_{\alpha}}{\Lambda^4}N_\mathrm{lin}^{\alpha} + \frac{f_{\alpha}^2}{\Lambda^8}N_\mathrm{quad}^{\alpha},
\end{equation}
where $N_\mathrm{SM}$ represents the contribution from SM processes, $N_\mathrm{lin}$ accounts for the interference between the EFT operator and the SM VBS processes, and $N_\mathrm{quad}$ corresponds to the pure contribution from the specific operator.
Unlike the approach outlined for the EW $\PZ\PV$ measurement, which fits the output distribution of a DNN optimized for the SM EW signature in the signal phase space, the invariant mass distributions of the diboson system, $M_{\PZ\PV}$ and $M_{\PW\PV}$, are used in the SRs for both $\PZ\PV$ and $\PW\PV$ channels. The $M_{\PW\PV}$ is defined as the invariant mass of the charged lepton, reconstructed neutrino and hadronically decaying vector boson. This observable is more sensitive to anomalous quartic gauge coupling effects~\cite{CMS:2019qfk}. 

Data yields from both the SRs and CRs are incorporated into the likelihood, as described in Section~\ref{sec:strategy} for the $\PZ\PV$ process and in Ref.~\cite{CMS:2021qzz} for the $\PW\PV$ process.

Systematic uncertainties are accounted for in the ML fit, as detailed in Section~\ref{sec:zvsystematics} for the $\PZ\PV$ channel and in Ref.~\cite{CMS:2021qzz} for the $\PW\PV$ channel. The primary sources of uncertainty in the measurement of the VBS $\PW\PV$ process include the statistical uncertainty of the data, the limited size of simulation samples, and uncertainties in background normalization. Among the theoretical uncertainties, the most significant arises from the choice of renormalization and factorization scales in the MC simulation of events. In the combination of the two channels, all background normalization uncertainties are uncorrelated, whereas the other experimental uncertainties are correlated.
The theoretical uncertainties associated with parton shower modeling and PDF uncertainties are also treated as correlated between the two channels. Similarly, each uncertainty related to the choice of renormalization and factorization scales in the MC simulation is treated as correlated when being applied across processes that are common to both channels.
	
\begin{figure}[htpb!]
    \centering
    \includegraphics[width=0.49\linewidth]{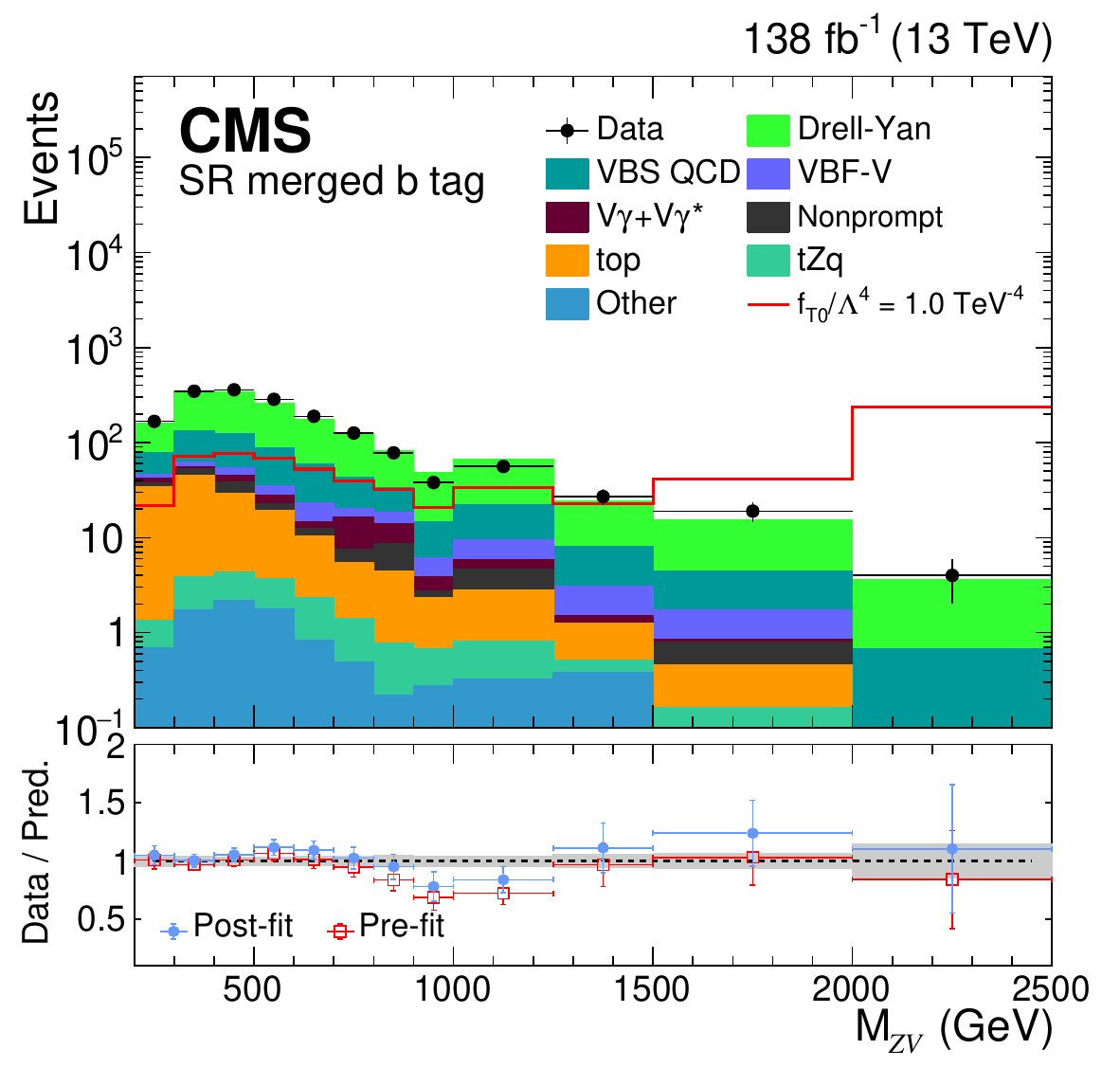}
    \includegraphics[width=0.49\linewidth]{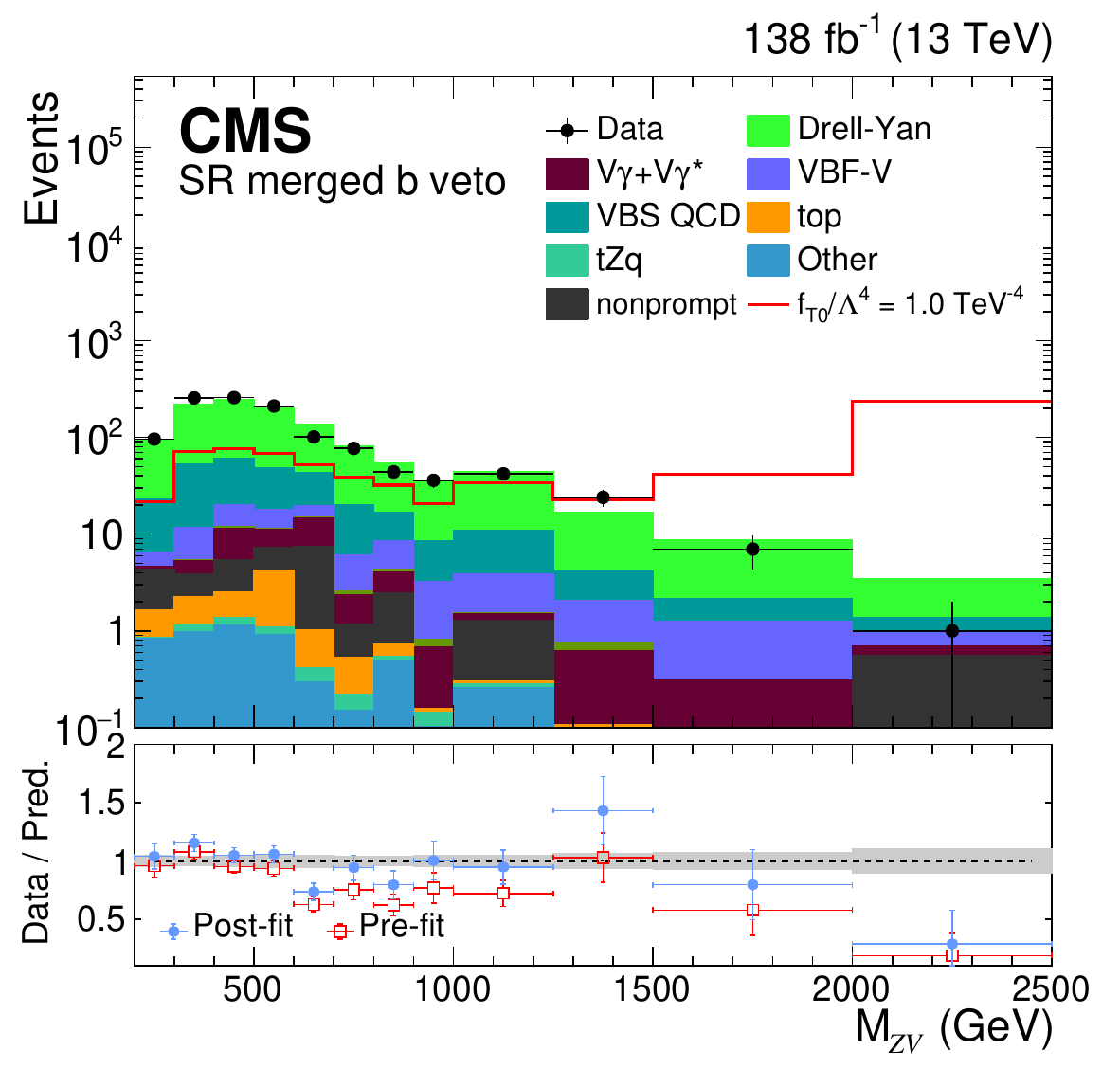}
    \caption{Distributions of $M_{\PZ\PV}$ for the data and post-fit backgrounds (stacked histograms), in the SRs of the merged \PQb tag (left) and the merged \PQb veto (right) channels. The template for one signal hypothesis is shown overlaid as a red solid line. The overflow is included in the last bin. The lower panels show the ratios of the data to the pre-fit background prediction and post-fit background yield as red open squares and blue points, respectively. The gray band in the lower panels indicates the systematic component of the post-fit background uncertainty. The vertical bars on the data points represent statistical uncertainties.}
 \label{fig:postfit_MVV_EFT_ZV_RunII}
\end{figure}

\begin{figure}[htpb!]
    \centering
    \includegraphics[width=0.49\linewidth]{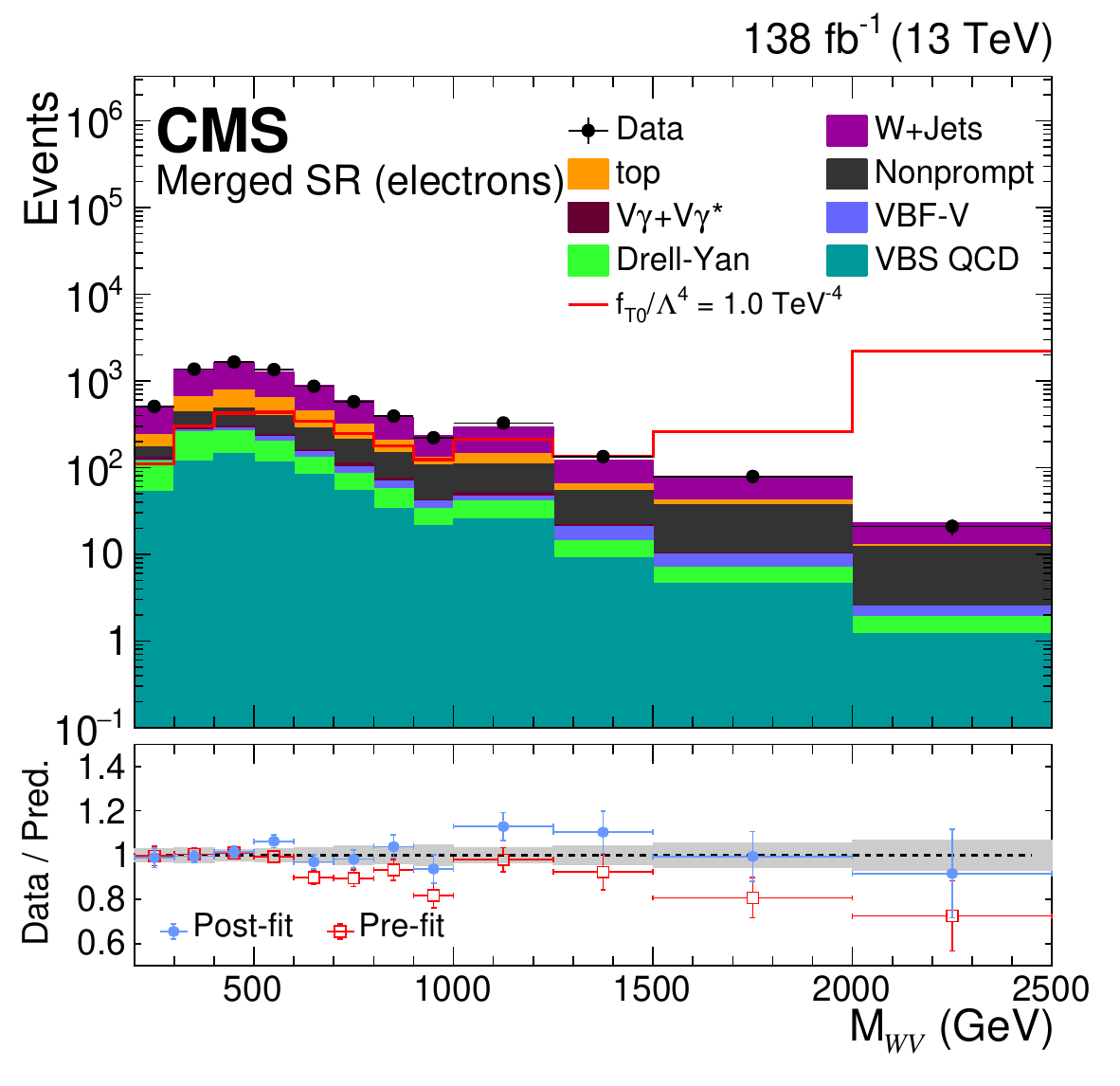}
    \includegraphics[width=0.49\linewidth]{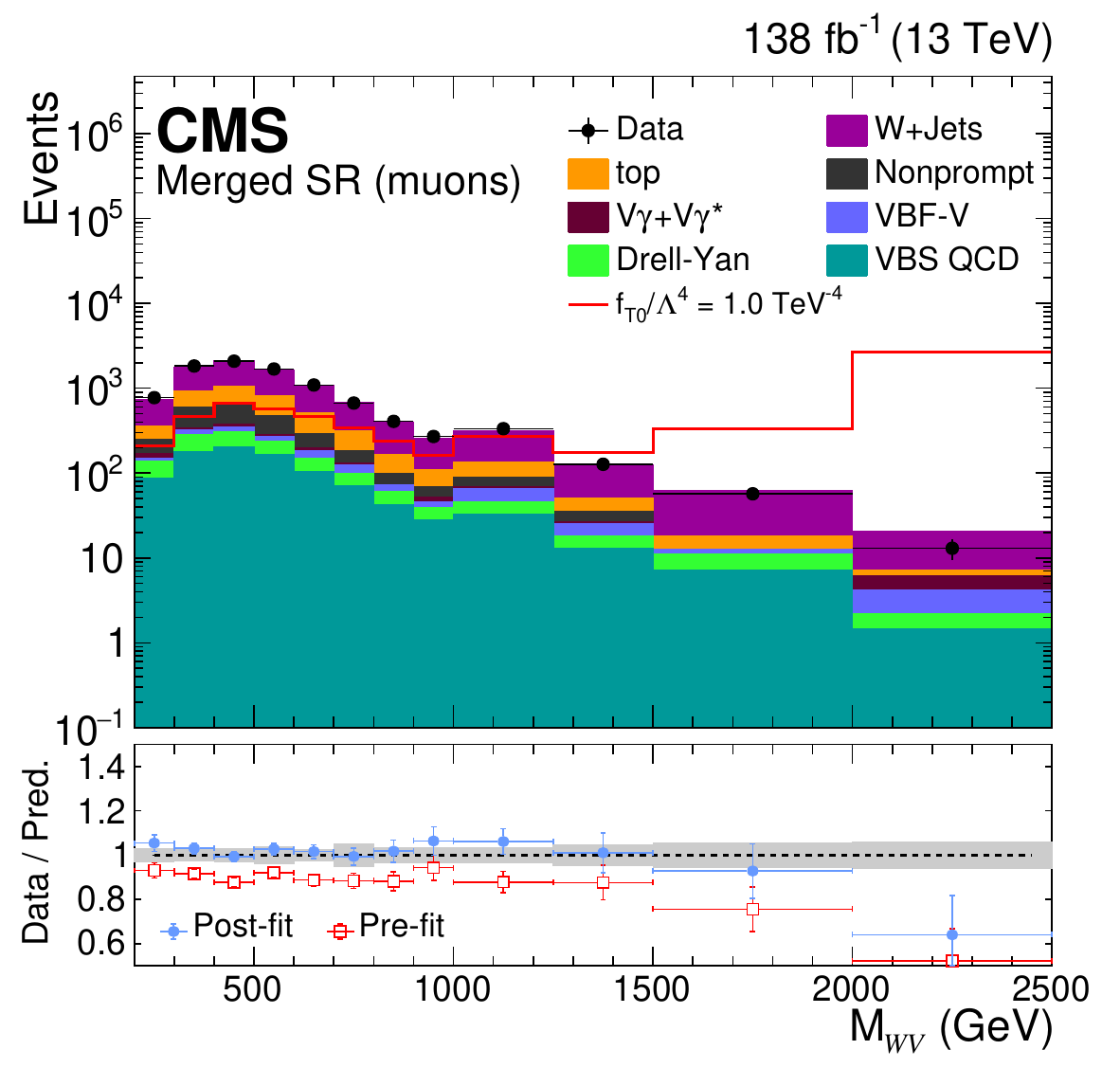}
    \caption{Distributions of $M_{\PW\PV}$ for the data and post-fit backgrounds (stacked histograms), in the SRs of the electron (left) and muon (right) channels in the merged regime. The template for one signal hypothesis is shown overlaid as a red solid line. The overflow is included in the last bin. The lower panels show the ratios of the data to the pre-fit background prediction and post-fit background yield as red open squares and blue points, respectively. The gray band in the lower panels indicates the systematic component of the post-fit background uncertainty. The vertical bars on the data points represent statistical uncertainties.}
 \label{fig:postfit_MVV_EFT_WV_RunII}
\end{figure}

No significant deviations from the SM predictions are observed, as can be seen in Figures~\ref{fig:postfit_MVV_EFT_ZV_RunII} and~\ref{fig:postfit_MVV_EFT_WV_RunII} for the $\PZ\PV$ and $\PW\PV$ channels, respectively.

\begin{figure}[htpb!]
    \centering
    \includegraphics[width=0.9\linewidth]{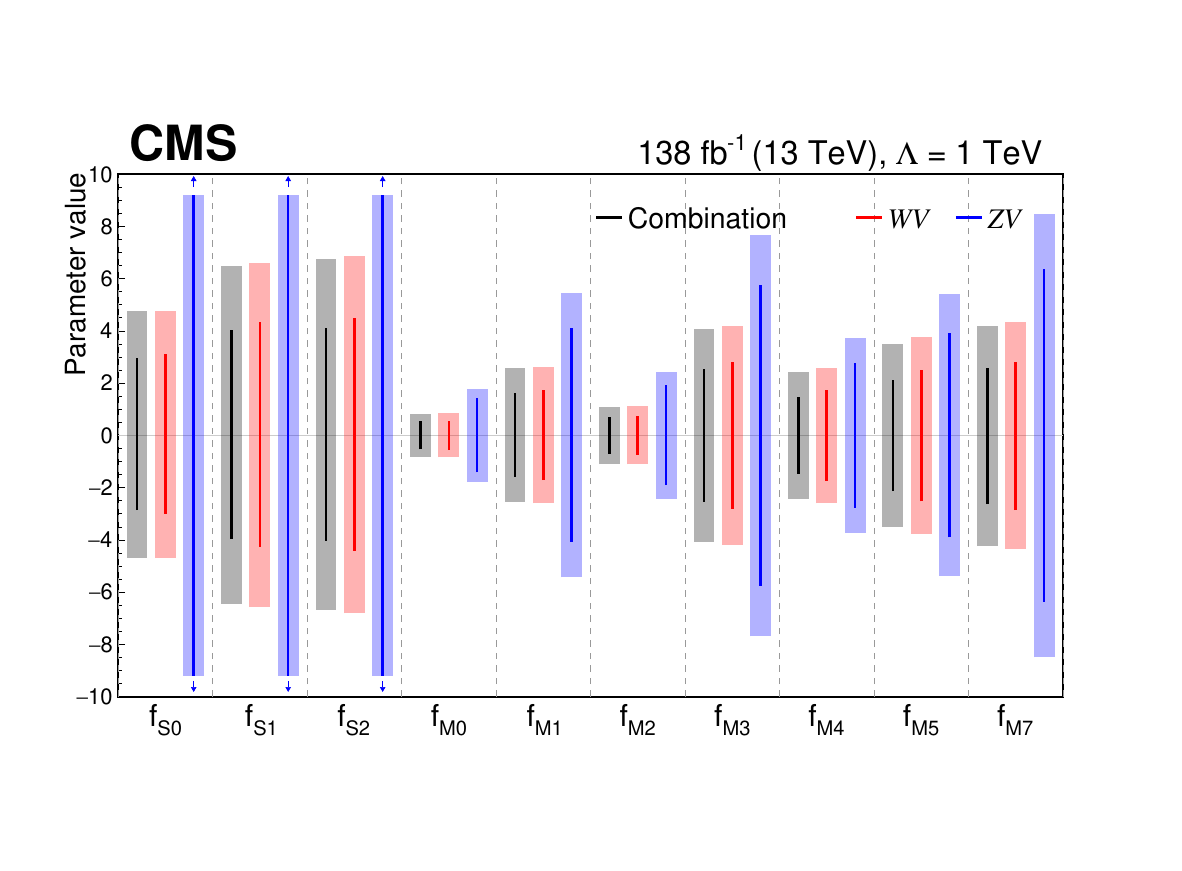} \\
    \includegraphics[width=0.9\linewidth]{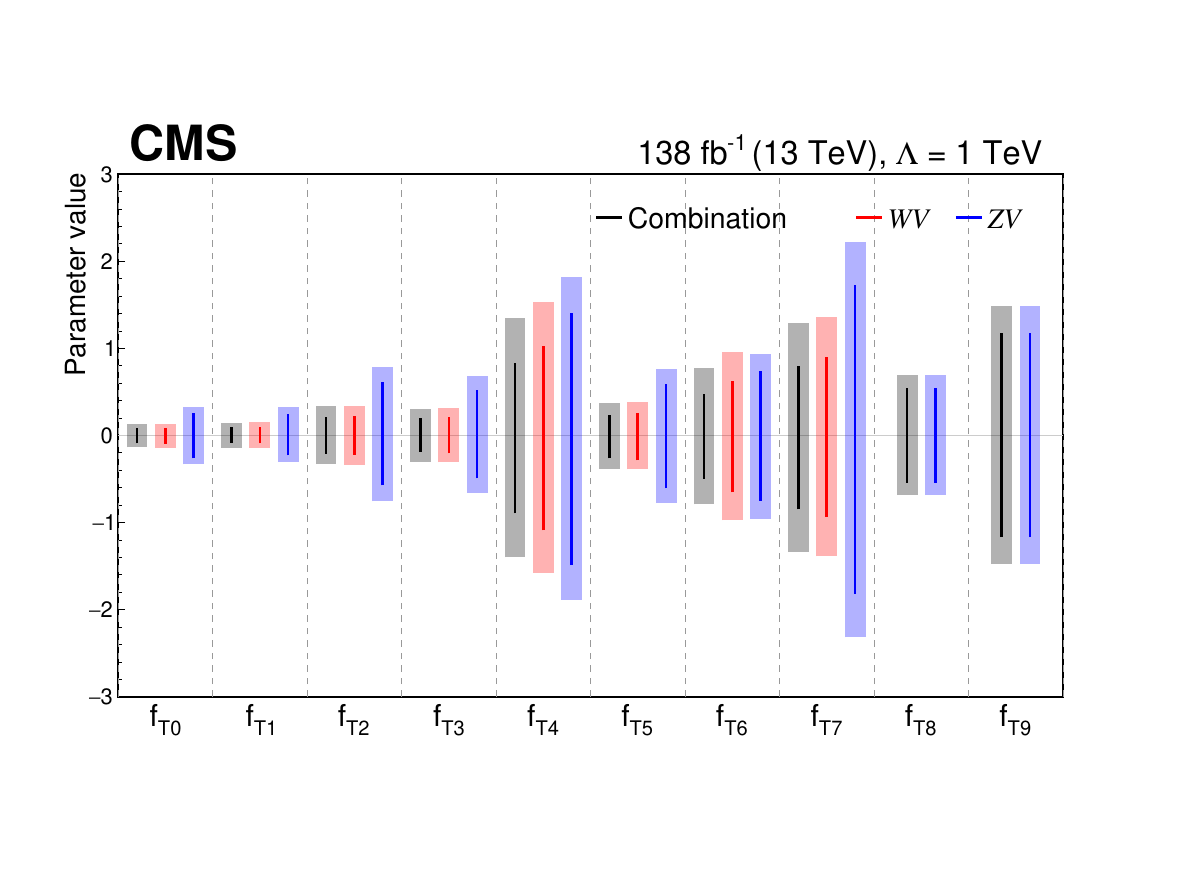}
    \caption{Constraints on dimension-8 Wilson coefficients derived from the $\PW\PV$, $\PZ\PV$, and their combined results. The shaded areas represent the expected 95\% \CL intervals, whereas the solid lines indicate the observed 95\% \CL limits. The top panel shows constraints on scalar and mixed operators, whereas the bottom panel shows transverse operators.}
    \label{fig:summary-cis}
\end{figure}

The 95\% confidence intervals on the Wilson coefficients are extracted through a likelihood scan performed by varying the corresponding Wilson coefficients one at a time and are reported in Table~\ref{tab:eftres} and in Figure~\ref{fig:summary-cis}. The signal strengths of the EW $\PZ\PV$ and $\PW\PV$ VBS processes are fixed to 1. To validate the results obtained from the asymptotic approximation, pseudo-experiments are generated for the signal and background, accounting for their statistical fluctuations. These results are consistent with those derived from Wilks' theorem~\cite{Bernlochner:2022oiw}.  
These results represent the best limits by CMS on these operators using the Run 2 data sample and represent the tightest world limits to date  using modern EFT modeling that properly includes systematic uncertainties in the EFT signals.

A direct comparison of these new results with an extrapolation of those from the previous CMS publication~\cite{CMS:2019qfk} is not possible, since no systematic uncertainties in the EFT signal were considered at the time. Results reported here show also that CMS limits are tighter than those reported by ATLAS on similar set of operators~\cite{ATLAS:2022nru} with a few exceptions (T5, T8 and T9). As a further cross-check, the obtained limits have been compared to unitarity bounds estimated using the formula in Ref.~\cite{Almeida:2020ylr}. The impact of applying a clipping unitarization scheme~\cite{CMS:2020fqz} at the corresponding unitarity bound was small, indicating that the presented limits are not significantly affected by unitarity considerations.

\begin{table*}[h!t]
\centering
\topcaption{
Observed and expected 95\% \CL intervals on the parameters of the quartic operators in the $\PW\PV$ and $\PZ\PV$ channels. The last two columns show the observed and expected limits for the combination of the $\PW\PV$ and $\PZ\PV$ channels. 
}
\label{tab:eftres}
\resizebox{\textwidth}{!}{
\begin{tabular}{*{7}{>{$}c<{$}}}
& \multicolumn{1}{c}{Observed ($\PZ\PV$)}  & \multicolumn{1}{c}{Expected ($\PZ\PV$)} & \multicolumn{1}{c}{Observed ($\PW\PV$)} & \multicolumn{1}{c}{Expected ($\PW\PV$)}& \multicolumn{1}{c}{Observed} & \multicolumn{1}{c}{Expected}  \\
& (\TeVns^{-4})   & (\TeVns^{-4})  & (\TeVns^{-4})   & (\TeVns^{-4}) & (\TeVns^{-4})   & (\TeVns^{-4})  \\
\hline
f_{\text{S0}} / \Lambda^4  & [-9.76,\, 9.89] & [-13.9,\, 14.0]& [-3.01,\, 3.1]    & [-4.7,\, 4.77]      & [-2.86,\, 2.96] & [-4.68,\, 4.75]\\
f_{\text{S1}} / \Lambda^4  & [-10.2,\, 10.3] & [-13.9,\, 13.9]& [-4.27,\, 4.32]   & [-6.56,\, 6.6]      & [-3.97,\, 4.02] & [-6.45,\, 6.49]\\
f_{\text{S2}} / \Lambda^4  & [-9.75,\, 9.89] & [-13.9,\, 14.0]& [-4.42,\, 4.48]   & [-6.81,\, 6.86]     & [-4.04,\, 4.11] & [-6.68,\, 6.73]\\
[\cmsTabSkip]
f_{\text{M0}} / \Lambda^4  & [-1.38,\, 1.38] & [-1.74,\, 1.74]& [-0.568,\, 0.567] & [-0.844,\, 0.843]   & [-0.539,\, 0.534] & [-0.828,\, 0.827]\\
f_{\text{M1}} / \Lambda^4  &  [-3.97,\, 4.00]& [-5.28,\, 5.29]& [-1.71,\, 1.75]   & [-2.6,\, 2.63]      & [-1.59,\, 1.62] & [-2.55,\, 2.58]\\
f_{\text{M2}} / \Lambda^4  &  [-1.86,\, 1.86]& [-2.37,\, 2.37]& [-0.746,\, 0.747] & [-1.11,\, 1.11]     & [-0.703,\, 0.703] & [-1.1,\, 1.1]\\
f_{\text{M3}} / \Lambda^4  &  [-5.60,\, 5.59]& [-7.47,\, 7.47]& [-2.81,\, 2.81]   & [-4.2,\, 4.2]       & [-2.55,\, 2.55] & [-4.08,\, 4.07]\\
f_{\text{M4}} / \Lambda^4  & [-2.70,\, 2.70] & [-3.61,\, 3.61]& [-1.74,\, 1.73]   & [-2.6,\, 2.59]      & [-1.48,\, 1.48] & [-2.42,\, 2.41]\\
f_{\text{M5}} / \Lambda^4  &  [-3.80,\, 3.81]& [-5.21,\, 5.23]& [-2.53,\, 2.51]   & [-3.77,\, 3.76]     & [-2.14,\, 2.13] & [-3.5,\, 3.5]\\
f_{\text{M7}} / \Lambda^4  &  [-6.09,\, 6.07]& [-8.26,\, 8.24]& [-2.86,\, 2.82]   & [-4.35,\, 4.32]     & [-2.63,\, 2.58] & [-4.24,\, 4.2]\\
[\cmsTabSkip]
f_{\text{T0}} / \Lambda^4  & [-0.26,\, 0.25] & [-0.33,\, 0.32]& [-0.096,\, 0.083] & [-0.14,\, 0.128]    & [-0.0921,\, 0.0785] & [-0.138,\, 0.127]\\
f_{\text{T1}} / \Lambda^4  &  [-0.22,\, 0.24]& [-0.30,\, 0.31]& [-0.0933,\, 0.1]  & [-0.142,\, 0.149]   & [-0.0863,\, 0.0943] & [-0.14,\, 0.147]\\
f_{\text{T2}} / \Lambda^4  &  [-0.56,\, 0.60]& [-0.74,\, 0.76]& [-0.225,\, 0.225] & [-0.336,\, 0.335]   &  [-0.21,\, 0.214] & [-0.331,\, 0.332]\\
f_{\text{T3}} / \Lambda^4  &  [-0.48,\, 0.51]& [-0.64,\, 0.66]& [-0.206,\, 0.206] & [-0.311,\, 0.31]    &  [-0.191,\, 0.194] & [-0.305,\, 0.305]\\
f_{\text{T4}} / \Lambda^4  &  [-1.44,\, 1.37]& [-1.84,\, 1.77]& [-1.09,\, 1.02]   & [-1.58,\, 1.53]     &  [-0.895,\, 0.828] & [-1.4,\, 1.35]\\
f_{\text{T5}} / \Lambda^4  &  [-0.59,\, 0.57]& [-0.76,\, 0.73]& [-0.287,\, 0.257] & [-0.391,\, 0.383]   &  [-0.265,\, 0.237] & [-0.382,\, 0.373]\\
f_{\text{T6}} / \Lambda^4  &  [-0.73,\, 0.71]& [-0.94,\, 0.92]& [-0.656,\, 0.627] & [-0.976,\, 0.954]   &  [-0.5,\, 0.478] & [-0.794,\, 0.775]\\
f_{\text{T7}} / \Lambda^4  &  [-1.78,\, 1.67]& [-2.26,\, 2.16]& [-0.936,\, 0.899] & [-1.39,\, 1.36]     &  [-0.85,\, 0.8] & [-1.34,\, 1.29]\\
f_{\text{T8}} / \Lambda^4  &  [-0.53,\, 0.53]& [-0.67,\, 0.67]& \NA             & \NA               &  [-0.53,\, 0.53]& [-0.67,\, 0.67]\\
f_{\text{T9}} / \Lambda^4  &  [-1.17,\, 1.16]& [-1.47,\, 1.45]& \NA             & \NA               &  [-1.17,\, 1.16]& [-1.47,\, 1.45]\\
\end{tabular}
}
\end{table*}

\section{Summary\label{sec:summary}}
A study was performed of the electroweak vector boson scattering production of $\PZ\PV$ ($\PV = \PW, \, \PZ$) boson pairs in lepton + jets decays in association with two forward jets in proton-proton collisions at 13\TeV. The analysis uses proton-proton collision events recorded by the CMS experiment at the LHC in 2016--2018, corresponding to an integrated luminosity of 138\fbinv.
The signal strength is measured in a region characterized by large invariant mass and pseudorapidity gap of the two forward jets using feed-forward deep neural network discriminators. The measured signal strength for electroweak $\PZ\PV$ vector boson scattering process is $0.63^{+0.53}_{-0.51}$.
Constraints are then established by combining $\PZ\PV$ and $\PW\PV$ data on coefficients of new operators from an effective field theory sensitive to an anomalous electroweak production of $\PW\PW$, $\PW\PZ$, and $\PZ\PZ$ boson pairs in association with two jets. Several world best limits are set on anomalous quartic gauge couplings in terms of dimension-8 standard model effective field theory operators. 

\begin{acknowledgments}
We congratulate our colleagues in the CERN accelerator departments for the excellent performance of the LHC and thank the technical and administrative staffs at CERN and at other CMS institutes for their contributions to the success of the CMS effort. In addition, we gratefully acknowledge the computing centers and personnel of the Worldwide LHC Computing Grid and other centers for delivering so effectively the computing infrastructure essential to our analyses. Finally, we acknowledge the enduring support for the construction and operation of the LHC, the CMS detector, and the supporting computing infrastructure provided by the following funding agencies: SC (Armenia), BMBWF and FWF (Austria); FNRS and FWO (Belgium); CNPq, CAPES, FAPERJ, FAPERGS, and FAPESP (Brazil); MES and BNSF (Bulgaria); CERN; CAS, MoST, and NSFC (China); MINCIENCIAS (Colombia); MSES and CSF (Croatia); RIF (Cyprus); SENESCYT (Ecuador); ERC PRG, TARISTU24-TK10 and MoER TK202 (Estonia); Academy of Finland, MEC, and HIP (Finland); CEA and CNRS/IN2P3 (France); SRNSF (Georgia); BMFTR, DFG, and HGF (Germany); GSRI (Greece); NKFIH (Hungary); DAE and DST (India); IPM (Iran); SFI (Ireland); INFN (Italy); MSIT and NRF (Republic of Korea); MES (Latvia); LMTLT (Lithuania); MOE and UM (Malaysia); BUAP, CINVESTAV, CONACYT, LNS, SEP, and UASLP-FAI (Mexico); MOS (Montenegro); MBIE (New Zealand); PAEC (Pakistan); MES, NSC, and NAWA (Poland); FCT (Portugal); MESTD (Serbia); MICIU/AEI and PCTI (Spain); MOSTR (Sri Lanka); Swiss Funding Agencies (Switzerland); MST (Taipei); MHESI (Thailand); TUBITAK and TENMAK (T\"{u}rkiye); NASU (Ukraine); STFC (United Kingdom); DOE and NSF (USA).

\hyphenation{Rachada-pisek} Individuals have received support from the Marie-Curie program and the European Research Council and Horizon 2020 Grant, contract Nos.\ 675440, 724704, 752730, 758316, 765710, 824093, 101115353, 101002207, 101001205, and COST Action CA16108 (European Union); the Leventis Foundation; the Alfred P.\ Sloan Foundation; the Alexander von Humboldt Foundation; the Science Committee, project no. 22rl-037 (Armenia); the Fonds pour la Formation \`a la Recherche dans l'Industrie et dans l'Agriculture (FRIA) and Fonds voor Wetenschappelijk Onderzoek contract No. 1228724N (Belgium); the Beijing Municipal Science \& Technology Commission, No. Z191100007219010, the Fundamental Research Funds for the Central Universities, the Ministry of Science and Technology of China under Grant No. 2023YFA1605804, and the Natural Science Foundation of China under Grant No. 12061141002 (China); the Ministry of Education, Youth and Sports (MEYS) of the Czech Republic; the Shota Rustaveli National Science Foundation, grant FR-22-985 (Georgia); the Deutsche Forschungsgemeinschaft (DFG), among others, under Germany's Excellence Strategy -- EXC 2121 ``Quantum Universe" -- 390833306, and under project number 400140256 - GRK2497; the Hellenic Foundation for Research and Innovation (HFRI), Project Number 2288 (Greece); the Hungarian Academy of Sciences, the New National Excellence Program - \'UNKP, the NKFIH research grants K 131991, K 133046, K 138136, K 143460, K 143477, K 146913, K 146914, K 147048, 2020-2.2.1-ED-2021-00181, TKP2021-NKTA-64, and 2021-4.1.2-NEMZ\_KI-2024-00036 (Hungary); the Council of Science and Industrial Research, India; ICSC -- National Research Center for High Performance Computing, Big Data and Quantum Computing, FAIR -- Future Artificial Intelligence Research, and CUP I53D23001070006 (Mission 4 Component 1), funded by the NextGenerationEU program (Italy); the Latvian Council of Science; the Ministry of Education and Science, project no. 2022/WK/14, and the National Science Center, contracts Opus 2021/41/B/ST2/01369, 2021/43/B/ST2/01552, 2023/49/B/ST2/03273, and the NAWA contract BPN/PPO/2021/1/00011 (Poland); the Funda\c{c}\~ao para a Ci\^encia e a Tecnologia, grant CEECIND/01334/2018 (Portugal); the National Priorities Research Program by Qatar National Research Fund; MICIU/AEI/10.13039/501100011033, ERDF/EU, "European Union NextGenerationEU/PRTR", and Programa Severo Ochoa del Principado de Asturias (Spain); the Chulalongkorn Academic into Its 2nd Century Project Advancement Project, the National Science, Research and Innovation Fund program IND\_FF\_68\_369\_2300\_097, and the Program Management Unit for Human Resources \& Institutional Development, Research and Innovation, grant B39G680009 (Thailand); the Kavli Foundation; the Nvidia Corporation; the SuperMicro Corporation; the Welch Foundation, contract C-1845; and the Weston Havens Foundation (USA).
\end{acknowledgments}

\bibliography{auto_generated}
\cleardoublepage \appendix\section{The CMS Collaboration \label{app:collab}}\begin{sloppypar}\hyphenpenalty=5000\widowpenalty=500\clubpenalty=5000\input{SMP-22-011-public-authorlist.tex}\end{sloppypar}
%%% END EDITABLE REGION %%%
% skeleton_end
\end{document}

%% file: SMP-22-011-public-authorlist.tex
\cmsinstitute{Yerevan Physics Institute, Yerevan, Armenia}
{\tolerance=6000
A.~Hayrapetyan, V.~Makarenko\cmsorcid{0000-0002-8406-8605}, A.~Tumasyan\cmsAuthorMark{1}\cmsorcid{0009-0000-0684-6742}
\par}
\cmsinstitute{Institut f\"{u}r Hochenergiephysik, Vienna, Austria}
{\tolerance=6000
W.~Adam\cmsorcid{0000-0001-9099-4341}, J.W.~Andrejkovic, L.~Benato\cmsorcid{0000-0001-5135-7489}, T.~Bergauer\cmsorcid{0000-0002-5786-0293}, K.~Damanakis\cmsorcid{0000-0001-5389-2872}, M.~Dragicevic\cmsorcid{0000-0003-1967-6783}, C.~Giordano, P.S.~Hussain\cmsorcid{0000-0002-4825-5278}, M.~Jeitler\cmsAuthorMark{2}\cmsorcid{0000-0002-5141-9560}, N.~Krammer\cmsorcid{0000-0002-0548-0985}, A.~Li\cmsorcid{0000-0002-4547-116X}, D.~Liko\cmsorcid{0000-0002-3380-473X}, I.~Mikulec\cmsorcid{0000-0003-0385-2746}, J.~Schieck\cmsAuthorMark{2}\cmsorcid{0000-0002-1058-8093}, R.~Sch\"{o}fbeck\cmsAuthorMark{2}\cmsorcid{0000-0002-2332-8784}, D.~Schwarz\cmsorcid{0000-0002-3821-7331}, M.~Shooshtari, M.~Sonawane\cmsorcid{0000-0003-0510-7010}, W.~Waltenberger\cmsorcid{0000-0002-6215-7228}, C.-E.~Wulz\cmsAuthorMark{2}\cmsorcid{0000-0001-9226-5812}
% ...
\par}
\vskip\cmsinstskip
\dag:~Deceased\\
$^{1}$Also at Yerevan State University, Yerevan, Armenia\\
$^{2}$Also at TU Wien, Vienna, Austria\\
% ...


%% file: SMP-22-011_temp.bbl
\providecommand{\href}[2]{#2}\begingroup\raggedright\begin{thebibliography}{10}%
\makeatletter
\providecommand{\hrefCMSnoop }[0]{\@secondoftwo}%
\makeatother
\providecommand{\doi}{\texttt{doi:}\begingroup \urlstyle{tt}\Url}

\bibitem{Green:2016trm}
\hrefCMSnoop {}{D.~R. Green, P.~Meade, and M.-A. Pleier, ``{Multiboson
  interactions at the LHC}'',} \textit{ Rev. Mod. Phys.} \textbf{ 89} (2017)
  035008,
  \href{http://dx.doi.org/10.1103/RevModPhys.89.035008}{\doi{10.1103/RevModPhys.89.035008}},
\href{http://www.arXiv.org/abs/1610.07572}{\texttt{arXiv:1610.07572}}.
%%CITATION = ARXIV:1610.07572;%%.

\bibitem{Anders:2018gfr}
\hrefCMSnoop {}{C.~F. Anders { et~al.}, ``{Vector boson scattering: Recent
  experimental and theory developments}'',} \textit{ Rev. Phys.} \textbf{ 3}
  (2018) 44,
  \href{http://dx.doi.org/10.1016/j.revip.2018.11.001}{\doi{10.1016/j.revip.2018.11.001}},
\href{http://www.arXiv.org/abs/1801.04203}{\texttt{arXiv:1801.04203}}.
%%CITATION = ARXIV:1801.04203;%%.

\bibitem{Aaboud:2016uuk}
\hrefCMSnoop {}{{ATLAS Collaboration}, ``{Search for anomalous electroweak
  production of $WW/WZ$ in association with a high-mass dijet system in pp
  collisions at $\sqrt{s}=8$ TeV with the ATLAS detector}'',} \textit{ Phys.
  Rev. D} \textbf{ 95} (2017) 032001,
  \href{http://dx.doi.org/10.1103/PhysRevD.95.032001}{\doi{10.1103/PhysRevD.95.032001}},
\href{http://www.arXiv.org/abs/1609.05122}{\texttt{arXiv:1609.05122}}.
%%CITATION = ARXIV:1609.05122;%%.

\bibitem{Sirunyan:2017fvv}
\hrefCMSnoop {}{{CMS Collaboration}, ``{Measurement of vector boson scattering
  and constraints on anomalous quartic couplings from events with four leptons
  and two jets in proton-proton collisions at $\sqrt{s}=$ 13 TeV}'',} \textit{
  Phys. Lett. B} \textbf{ 774} (2017) 682,
  \href{http://dx.doi.org/10.1016/j.physletb.2017.10.020}{\doi{10.1016/j.physletb.2017.10.020}},
\href{http://www.arXiv.org/abs/1708.02812}{\texttt{arXiv:1708.02812}}.
%%CITATION = ARXIV:1708.02812;%%.

\bibitem{ATLAS:2022xyx}
\href {https://cds.cern.ch/record/2816369}{{ATLAS Collaboration}, ``{Combined
  effective field theory interpretation of Higgs boson and weak boson
  production and decay with ATLAS data and electroweak precision
  observables}'',} technical report, CERN, 2022.

\bibitem{ATLAS:2024jpa}
\href {https://cds.cern.ch/record/2896020}{{ATLAS Collaboration}, ``{Top EFT
  summary plots April 2024}'',} technical report, CERN, 2024.

\bibitem{Covarelli:2021gyz}
\hrefCMSnoop {}{R.~Covarelli, M.~Pellen, and M.~Zaro, ``{Vector-boson
  scattering at the LHC: Unraveling the electroweak sector}'',} \textit{ Int.
  J. Mod. Phys. A} \textbf{ 36} (2021) 2130009,
  \href{http://dx.doi.org/10.1142/S0217751X2130009X}{\doi{10.1142/S0217751X2130009X}},
  \href{http://www.arXiv.org/abs/2102.10991}{\texttt{arXiv:2102.10991}}.

\bibitem{CMS:2021cxr}
\hrefCMSnoop {}{{CMS Collaboration}, ``{Measurement of W$^\pm\gamma$
  differential cross sections in proton-proton collisions at $\sqrt{s}$ = 13
  TeV and effective field theory constraints}'',} \textit{ Phys. Rev. D}
  \textbf{ 105} (2022) 052003,
  \href{http://dx.doi.org/10.1103/PhysRevD.105.052003}{\doi{10.1103/PhysRevD.105.052003}},
  \href{http://www.arXiv.org/abs/2111.13948}{\texttt{arXiv:2111.13948}}.

\bibitem{CMS:2024skm}
\hrefCMSnoop {}{{CMS Collaboration}, ``{Observation of
  $\gamma\gamma\to\tau\tau$ in proton-proton collisions and limits on the
  anomalous electromagnetic moments of the $\tau$ lepton}'',} \textit{ Rept.
  Prog. Phys.} \textbf{ 87} (2024) 107801,
  \href{http://dx.doi.org/10.1088/1361-6633/ad6fcb}{\doi{10.1088/1361-6633/ad6fcb}},
  \href{http://www.arXiv.org/abs/2406.03975}{\texttt{arXiv:2406.03975}}.

\bibitem{ATLAS:2024ett}
\hrefCMSnoop {}{{ATLAS Collaboration}, ``{Observation of electroweak production
  of W$^{+}$W$^{-}$ in association with jets in proton-proton collisions at $
  \sqrt{s} =$ 13 TeV with the ATLAS detector}'',} \textit{ JHEP} \textbf{ 07}
  (2024) 254,
  \href{http://dx.doi.org/10.1007/JHEP07(2024)254}{\doi{10.1007/JHEP07(2024)254}},
  \href{http://www.arXiv.org/abs/2403.04869}{\texttt{arXiv:2403.04869}}.

\bibitem{ATLAS:2018mxa}
\hrefCMSnoop {}{{ATLAS Collaboration}, ``{Observation of electroweak $W^{\pm}Z$
  boson pair production in association with two jets in $pp$ collisions at
  $\sqrt{s} =$ 13 TeV with the ATLAS detector}'',} \textit{ Phys. Lett. B}
  \textbf{ 793} (2019) 469,
  \href{http://dx.doi.org/10.1016/j.physletb.2019.05.012}{\doi{10.1016/j.physletb.2019.05.012}},
  \href{http://www.arXiv.org/abs/1812.09740}{\texttt{arXiv:1812.09740}}.

\bibitem{CMS:2020etf}
\hrefCMSnoop {}{{CMS Collaboration}, ``{Measurements of production cross
  sections of polarized same-sign W boson pairs in association with two jets in
  proton-proton collisions at $\sqrt{s} =$ 13 TeV}'',} \textit{ Phys. Lett. B}
  \textbf{ 812} (2021) 136018,
  \href{http://dx.doi.org/10.1016/j.physletb.2020.136018}{\doi{10.1016/j.physletb.2020.136018}},
  \href{http://www.arXiv.org/abs/2009.09429}{\texttt{arXiv:2009.09429}}.

\bibitem{CMS:2020fqz}
\hrefCMSnoop {}{{CMS Collaboration}, ``{Evidence for electroweak production of
  four charged leptons and two jets in proton-proton collisions at $\sqrt {s}$
  = 13 TeV}'',} \textit{ Phys. Lett. B} \textbf{ 812} (2021) 135992,
  \href{http://dx.doi.org/10.1016/j.physletb.2020.135992}{\doi{10.1016/j.physletb.2020.135992}},
  \href{http://www.arXiv.org/abs/2008.07013}{\texttt{arXiv:2008.07013}}.

\bibitem{CMS:2020gfh}
\hrefCMSnoop {}{{CMS Collaboration}, ``{Measurements of production cross
  sections of WZ and same-sign WW boson pairs in association with two jets in
  proton-proton collisions at $\sqrt{s} =$ 13 TeV}'',} \textit{ Phys. Lett. B}
  \textbf{ 809} (2020) 135710,
  \href{http://dx.doi.org/10.1016/j.physletb.2020.135710}{\doi{10.1016/j.physletb.2020.135710}},
  \href{http://www.arXiv.org/abs/2005.01173}{\texttt{arXiv:2005.01173}}.

\bibitem{ATLAS:2020nlt}
\hrefCMSnoop {}{{ATLAS Collaboration}, ``{Observation of electroweak production
  of two jets and a {Z}-boson pair}'',} \textit{ Nature Phys.} \textbf{ 19}
  (2023) 237,
  \href{http://dx.doi.org/10.1038/s41567-022-01757-y}{\doi{10.1038/s41567-022-01757-y}},
  \href{http://www.arXiv.org/abs/2004.10612}{\texttt{arXiv:2004.10612}}.

\bibitem{CMS:2022woe}
\hrefCMSnoop {}{{CMS Collaboration}, ``{Observation of electroweak W$^+$W$^-$
  pair production in association with two jets in proton-proton collisions at
  $\sqrt{s}=13$ TeV}'',} \textit{ Phys. Lett. B} \textbf{ 841} (2023) 137495,
  \href{http://dx.doi.org/10.1016/j.physletb.2022.137495}{\doi{10.1016/j.physletb.2022.137495}},
  \href{http://www.arXiv.org/abs/2205.05711}{\texttt{arXiv:2205.05711}}.

\bibitem{ATLAS:2023sua}
\hrefCMSnoop {}{{ATLAS Collaboration}, ``{Measurement and interpretation of
  same-sign W boson pair production in association with two jets in pp
  collisions at $ \sqrt{s} =$ 13 TeV with the ATLAS detector}'',} \textit{
  JHEP} \textbf{ 04} (2024) 026,
  \href{http://dx.doi.org/10.1007/JHEP04(2024)026}{\doi{10.1007/JHEP04(2024)026}},
  \href{http://www.arXiv.org/abs/2312.00420}{\texttt{arXiv:2312.00420}}.

\bibitem{SSWW-CMS}
\hrefCMSnoop {}{{CMS Collaboration}, ``{Observation of electroweak production
  of same-sign $W$ boson pairs in the two jet and two same-sign lepton final
  state in proton-proton collisions at $\sqrt{s} = $ 13 TeV}'',} \textit{ Phys.
  Rev. Lett.} \textbf{ 120} (2018) 081801,
  \href{http://dx.doi.org/10.1103/PhysRevLett.120.081801}{\doi{10.1103/PhysRevLett.120.081801}},
  \href{http://www.arXiv.org/abs/1709.05822}{\texttt{arXiv:1709.05822}}.

\bibitem{SSWW-ATLAS}
\hrefCMSnoop {}{{ATLAS Collaboration}, ``{Observation of electroweak production
  of a same-sign $W$ boson pair in association with two jets in pp collisions
  at $\sqrt{s}=13$ TeV with the ATLAS detector}'',} \textit{ Phys. Rev. Lett.}
  \textbf{ 123} (2019) 161801,
  \href{http://dx.doi.org/10.1103/PhysRevLett.123.161801}{\doi{10.1103/PhysRevLett.123.161801}},
  \href{http://www.arXiv.org/abs/1906.03203}{\texttt{arXiv:1906.03203}}.

\bibitem{ATLAS:2019thr}
\hrefCMSnoop {}{{ATLAS Collaboration}, ``{Search for the electroweak diboson
  production in association with a high-mass dijet system in semileptonic final
  states in pp collisions at $\sqrt{s}=13$ TeV with the ATLAS detector}'',}
  \textit{ Phys. Rev. D} \textbf{ 100} (2019) 032007,
  \href{http://dx.doi.org/10.1103/PhysRevD.100.032007}{\doi{10.1103/PhysRevD.100.032007}},
  \href{http://www.arXiv.org/abs/1905.07714}{\texttt{arXiv:1905.07714}}.

\bibitem{CMS:2019qfk}
\hrefCMSnoop {}{{CMS Collaboration}, ``{Search for anomalous electroweak
  production of vector boson pairs in association with two jets in
  proton-proton collisions at 13 TeV}'',} \textit{ Phys. Lett. B} \textbf{ 798}
  (2019) 134985,
  \href{http://dx.doi.org/10.1016/j.physletb.2019.134985}{\doi{10.1016/j.physletb.2019.134985}},
  \href{http://www.arXiv.org/abs/1905.07445}{\texttt{arXiv:1905.07445}}.

\bibitem{CMS:2021qzz}
\hrefCMSnoop {}{{CMS Collaboration}, ``{Evidence for WW/WZ vector boson
  scattering in the decay channel \ensuremath{\ell}\ensuremath{\nu}qq produced
  in association with two jets in proton-proton collisions at $\sqrt{s}=13$
  TeV}'',} \textit{ Phys. Lett. B} \textbf{ 834} (2022) 137438,
  \href{http://dx.doi.org/10.1016/j.physletb.2022.137438}{\doi{10.1016/j.physletb.2022.137438}},
  \href{http://www.arXiv.org/abs/2112.05259}{\texttt{arXiv:2112.05259}}.

\bibitem{BuchmuellerWyler}
\hrefCMSnoop {}{W.~Buchm{\"{u}}ller and D.~Wyler, ``{Effective lagrangian
  analysis of new interactions and flavour conservation}'',} \textit{ Nucl.
  Phys. B} \textbf{ 268} (1986) 621,
  \href{http://dx.doi.org/10.1016/0550-3213(86)90262-2}{\doi{10.1016/0550-3213(86)90262-2}}.

\bibitem{warsaw}
\hrefCMSnoop {}{B.~Grzadkowski, M.~Iskrzy{\'n}ski, M.~Misiak, and J.~Rosiek,
  ``{Dimension-six terms in the Standard Model Lagrangian}'',} \textit{ JHEP}
  \textbf{ 10} (2010) 085,
  \href{http://dx.doi.org/10.1007/JHEP10(2010)085}{\doi{10.1007/JHEP10(2010)085}},
  \href{http://www.arXiv.org/abs/1008.4884}{\texttt{arXiv:1008.4884}}.

\bibitem{PhysRevD.93.093013}
\hrefCMSnoop {}{O.~J.~P. {\'E}boli and M.~C. Gonzalez-Garcia, ``{Classifying
  the bosonic quartic couplings}'',} \textit{ Phys. Rev. D} \textbf{ 93} (2016)
  093013,
  \href{http://dx.doi.org/10.1103/PhysRevD.93.093013}{\doi{10.1103/PhysRevD.93.093013}},
  \href{http://www.arXiv.org/abs/1604.03555}{\texttt{arXiv:1604.03555}}.

\bibitem{Falkowski:2016cxu}
A.~Falkowski\hrefCMSnoop {}{ { et~al.}, ``{Anomalous triple gauge couplings in
  the effective field theory approach at the LHC}'',} \textit{ JHEP} \textbf{
  02} (2017) 115,
  \href{http://dx.doi.org/10.1007/JHEP02(2017)115}{\doi{10.1007/JHEP02(2017)115}},
  \href{http://www.arXiv.org/abs/1609.06312}{\texttt{arXiv:1609.06312}}.

\bibitem{Durieux:2024zrg}
G.~Durieux\hrefCMSnoop {}{ { et~al.}, ``{LHC EFT WG note: Basis for anomalous
  quartic gauge couplings}'',} \textit{ SciPost Phys. Comm. Rep.} (11, 2024)
  \href{http://dx.doi.org/10.21468/SciPostPhysCommRep.6}{\doi{10.21468/SciPostPhysCommRep.6}},
  \href{http://www.arXiv.org/abs/2411.02483}{\texttt{arXiv:2411.02483}}.

\bibitem{Almeida:2020ylr}
\hrefCMSnoop {}{E.~d.~S. Almeida, O.~J.~P. {\'E}boli, and M.~C.
  Gonzalez\textendash{}Garcia, ``{Unitarity constraints on anomalous quartic
  couplings}'',} \textit{ Phys. Rev. D} \textbf{ 101} (2020) 113003,
  \href{http://dx.doi.org/10.1103/PhysRevD.101.113003}{\doi{10.1103/PhysRevD.101.113003}},
  \href{http://www.arXiv.org/abs/2004.05174}{\texttt{arXiv:2004.05174}}.

\bibitem{hepdata}
\hrefCMSnoop {}{}{HEPData} record for this analysis, 2025.
\newblock
  \href{http://dx.doi.org/10.17182/hepdata.165281}{\doi{10.17182/hepdata.165281}}.

\bibitem{CMS:2008xjf}
\hrefCMSnoop {}{{CMS Collaboration}, ``The {CMS} experiment at the {CERN}
  {LHC}'',} \textit{ JINST} \textbf{ 3} (2008) 8004,
  \href{http://dx.doi.org/10.1088/1748-0221/3/08/S08004}{\doi{10.1088/1748-0221/3/08/S08004}}.

\bibitem{CMS:2023gfb}
\hrefCMSnoop {}{{CMS Collaboration}, ``Development of the {CMS} detector for
  the {CERN LHC Run 3}'',} \textit{ JINST} \textbf{ 19} (2024) P05064,
  \href{http://dx.doi.org/10.1088/1748-0221/19/05/P05064}{\doi{10.1088/1748-0221/19/05/P05064}},
  \href{http://www.arXiv.org/abs/2309.05466}{\texttt{arXiv:2309.05466}}.

\bibitem{CMS:2020cmk}
\hrefCMSnoop {}{{CMS Collaboration}, ``{Performance of the CMS Level-1 trigger
  in proton-proton collisions at $\sqrt{s} = 13$\,TeV}'',} \textit{ JINST}
  \textbf{ 15} (2020) P10017,
  \href{http://dx.doi.org/10.1088/1748-0221/15/10/P10017}{\doi{10.1088/1748-0221/15/10/P10017}},
  \href{http://www.arXiv.org/abs/2006.10165}{\texttt{arXiv:2006.10165}}.

\bibitem{CMS:2016ngn}
\hrefCMSnoop {}{{CMS Collaboration}, ``{The CMS trigger system}'',} \textit{
  JINST} \textbf{ 12} (2017) P01020,
  \href{http://dx.doi.org/10.1088/1748-0221/12/01/P01020}{\doi{10.1088/1748-0221/12/01/P01020}},
\href{http://www.arXiv.org/abs/1609.02366}{\texttt{arXiv:1609.02366}}.
%%CITATION = ARXIV:1609.02366;%%.

\bibitem{CMS:2024aqx}
\hrefCMSnoop {}{{CMS Collaboration}, ``Performance of the {CMS} high-level
  trigger during {LHC} {Run} 2'',} \textit{ JINST} \textbf{ 19} (2024) P11021,
  \href{http://dx.doi.org/10.1088/1748-0221/19/11/P11021}{\doi{10.1088/1748-0221/19/11/P11021}},
  \href{http://www.arXiv.org/abs/2410.17038}{\texttt{arXiv:2410.17038}}.

\bibitem{CMS:2020uim}
\hrefCMSnoop {}{{CMS Collaboration}, ``Electron and photon reconstruction and
  identification with the {CMS} experiment at the {CERN} {LHC}'',} \textit{
  JINST} \textbf{ 16} (2021) P05014,
  \href{http://dx.doi.org/10.1088/1748-0221/16/05/P05014}{\doi{10.1088/1748-0221/16/05/P05014}},
  \href{http://www.arXiv.org/abs/2012.06888}{\texttt{arXiv:2012.06888}}.

\bibitem{CMS:2018rym}
\hrefCMSnoop {}{{CMS Collaboration}, ``{Performance of the CMS muon detector
  and muon reconstruction with proton-proton collisions at $\sqrt{s}=$ 13
  TeV}'',} \textit{ JINST} \textbf{ 13} (2018) P06015,
  \href{http://dx.doi.org/10.1088/1748-0221/13/06/P06015}{\doi{10.1088/1748-0221/13/06/P06015}},
  \href{http://www.arXiv.org/abs/1804.04528}{\texttt{arXiv:1804.04528}}.

\bibitem{CMS:2014pgm}
\hrefCMSnoop {}{{CMS Collaboration}, ``{Description and performance of track
  and primary-vertex reconstruction with the CMS tracker}'',} \textit{ JINST}
  \textbf{ 9} (2014) 10009,
  \href{http://dx.doi.org/10.1088/1748-0221/9/10/P10009}{\doi{10.1088/1748-0221/9/10/P10009}},
  \href{http://www.arXiv.org/abs/1405.6569}{\texttt{arXiv:1405.6569}}.

\bibitem{Alwall:2014hca}
J.~Alwall\hrefCMSnoop {}{ { et~al.}, ``{The automated computation of tree-level
  and next-to-leading order differential cross sections, and their matching to
  parton shower simulations}'',} \textit{ JHEP} \textbf{ 07} (2014) 079,
  \href{http://dx.doi.org/10.1007/JHEP07(2014)079}{\doi{10.1007/JHEP07(2014)079}},
  \href{http://www.arXiv.org/abs/1405.0301}{\texttt{arXiv:1405.0301}}.

\bibitem{reweight2016}
\hrefCMSnoop {}{O.~Mattelaer, ``{On the maximal use of Monte Carlo samples:
  reweighting events at NLO accuracy}'',} \textit{ Eur. Phys. J. C} \textbf{
  76} (2016) 674,
  \href{http://dx.doi.org/10.1140/epjc/s10052-016-4533-7}{\doi{10.1140/epjc/s10052-016-4533-7}},
  \href{http://www.arXiv.org/abs/1607.00763}{\texttt{arXiv:1607.00763}}.

\bibitem{Belvedere:2024nzh}
\hrefCMSnoop {}{A.~Belvedere { et~al.}, ``{LHC EFT WG note: SMEFT predictions,
  event reweighting, and simulation}'',} \textit{ SciPost Phys. Comm. Rep.}
  ({2024}) 4,
  \href{http://dx.doi.org/10.21468/SciPostPhysCommRep.4}{\doi{10.21468/SciPostPhysCommRep.4}},
  \href{http://www.arXiv.org/abs/2406.14620}{\texttt{arXiv:2406.14620}}.

\bibitem{Frederix:2012ps}
\hrefCMSnoop {}{R.~Frederix and S.~Frixione, ``{Merging meets matching in
  MC@NLO}'',} \textit{ JHEP} \textbf{ 12} (2012) 061,
  \href{http://dx.doi.org/10.1007/JHEP12(2012)061}{\doi{10.1007/JHEP12(2012)061}},
  \href{http://www.arXiv.org/abs/1209.6215}{\texttt{arXiv:1209.6215}}.

\bibitem{Alioli:2010xd}
\hrefCMSnoop {}{S.~Alioli, P.~Nason, C.~Oleari, and E.~Re, ``{A general
  framework for implementing NLO calculations in shower Monte Carlo programs:
  the POWHEG BOX}'',} \textit{ JHEP} \textbf{ 06} (2010) 043,
  \href{http://dx.doi.org/10.1007/JHEP06(2010)043}{\doi{10.1007/JHEP06(2010)043}},
\href{http://www.arXiv.org/abs/1002.2581}{\texttt{arXiv:1002.2581}}.
%%CITATION = ARXIV:1002.2581;%%.

\bibitem{Czakon:2017wor}
M.~Czakon\hrefCMSnoop {}{ { et~al.}, ``{Top-pair production at the LHC through
  NNLO QCD and NLO EW}'',} \textit{ JHEP} \textbf{ 10} (2017) 186,
  \href{http://dx.doi.org/10.1007/JHEP10(2017)186}{\doi{10.1007/JHEP10(2017)186}},
  \href{http://www.arXiv.org/abs/1705.04105}{\texttt{arXiv:1705.04105}}.

\bibitem{CMS:2016oae}
\hrefCMSnoop {}{{CMS Collaboration}, ``{Measurement of differential cross
  sections for top quark pair production using the lepton + jets final state in
  proton-proton collisions at 13\TeV}'',} \textit{ Phys. Rev. D} \textbf{ 95}
  (2017) 092001,
  \href{http://dx.doi.org/10.1103/PhysRevD.95.092001}{\doi{10.1103/PhysRevD.95.092001}},
  \href{http://www.arXiv.org/abs/1610.04191}{\texttt{arXiv:1610.04191}}.

\bibitem{CMS:2019raw}
\hrefCMSnoop {}{{CMS Collaboration}, ``{Measurements of differential $\PZ$
  boson production cross sections in proton-proton collisions at $\sqrt{s} =
  13\TeV$}'',} \textit{ JHEP} \textbf{ 12} (2019) 061,
  \href{http://dx.doi.org/10.1007/JHEP12(2019)061}{\doi{10.1007/JHEP12(2019)061}},
  \href{http://www.arXiv.org/abs/1909.04133}{\texttt{arXiv:1909.04133}}.

\bibitem{Nason:2013ydw}
\hrefCMSnoop {}{P.~Nason and G.~Zanderighi, ``{$W^+ W^-$ , $W Z$ and $Z Z$
  production in the POWHEG-BOX-V2}'',} \textit{ Eur. Phys. J. C} \textbf{ 74}
  (2014) 2702,
  \href{http://dx.doi.org/10.1140/epjc/s10052-013-2702-5}{\doi{10.1140/epjc/s10052-013-2702-5}},
  \href{http://www.arXiv.org/abs/1311.1365}{\texttt{arXiv:1311.1365}}.

\bibitem{Ball_2015}
\hrefCMSnoop {}{{NNPDF} Collaboration, ``{Parton distributions for the LHC Run
  II}'',} \textit{ JHEP} \textbf{ 04} (2015) 040,
  \href{http://dx.doi.org/10.1007/JHEP04(2015)040}{\doi{10.1007/JHEP04(2015)040}},
  \href{http://www.arXiv.org/abs/1410.8849}{\texttt{arXiv:1410.8849}}.

\bibitem{thennpdfcollaboration2017parton}
\hrefCMSnoop {}{{NNPDF} Collaboration, ``{Parton distributions from
  high-precision collider data}'',} \textit{ Eur. Phys. J. C} \textbf{ 77}
  (2017) 663,
  \href{http://dx.doi.org/10.1140/epjc/s10052-017-5199-5}{\doi{10.1140/epjc/s10052-017-5199-5}},
  \href{http://www.arXiv.org/abs/1706.00428}{\texttt{arXiv:1706.00428}}.

\bibitem{Sjostrand:2014zea}
T.~Sj{\"o}strand\hrefCMSnoop {}{ { et~al.}, ``{An Introduction to PYTHIA
  8.2}'',} \textit{ Comput. Phys. Commun.} \textbf{ 191} (2015) 159,
  \href{http://dx.doi.org/10.1016/j.cpc.2015.01.024}{\doi{10.1016/j.cpc.2015.01.024}},
\href{http://www.arXiv.org/abs/1410.3012}{\texttt{arXiv:1410.3012}}.
%%CITATION = ARXIV:1410.3012;%%.

\bibitem{Khachatryan_2016}
\hrefCMSnoop {}{{CMS Collaboration}, ``{Event generator tunes obtained from
  underlying event and multiparton scattering measurements}'',} \textit{ Eur.
  Phys. J. C} \textbf{ 76} (2016) 155,
  \href{http://dx.doi.org/10.1140/epjc/s10052-016-3988-x}{\doi{10.1140/epjc/s10052-016-3988-x}},
  \href{http://www.arXiv.org/abs/1512.00815}{\texttt{arXiv:1512.00815}}.

\bibitem{Sirunyan_2020}
\hrefCMSnoop {}{{CMS Collaboration}, ``{Extraction and validation of a new set
  of CMS PYTHIA8 tunes from underlying-event measurements}'',} \textit{ Eur.
  Phys. J. C} \textbf{ 80} (2020) 4,
  \href{http://dx.doi.org/10.1140/epjc/s10052-019-7499-4}{\doi{10.1140/epjc/s10052-019-7499-4}},
  \href{http://www.arXiv.org/abs/1903.12179}{\texttt{arXiv:1903.12179}}.

\bibitem{ballestrero_precise_2018}
\hrefCMSnoop {}{A.~Ballestrero { et~al.}, ``{Precise predictions for same-sign
  $\PW$-boson scattering at the LHC}'',} \textit{ Eur. Phys. J. C} \textbf{ 78}
  (2018) 671,
  \href{http://dx.doi.org/10.1140/epjc/s10052-018-6136-y}{\doi{10.1140/epjc/s10052-018-6136-y}},
  \href{http://www.arXiv.org/abs/1803.07943}{\texttt{arXiv:1803.07943}}.

\bibitem{Jager:2020hkz}
B.~J{\"a}ger\hrefCMSnoop {}{ { et~al.}, ``{Parton-shower effects in Higgs
  production via vector-boson fusion}'',} \textit{ Eur. Phys. J. C} \textbf{
  80} (2020) 756,
  \href{http://dx.doi.org/10.1140/epjc/s10052-020-8326-7}{\doi{10.1140/epjc/s10052-020-8326-7}},
  \href{http://www.arXiv.org/abs/2003.12435}{\texttt{arXiv:2003.12435}}.

\bibitem{Agostinelli:2002hh}
\hrefCMSnoop {}{{GEANT4} Collaboration, ``{GEANT4: A simulation toolkit}'',}
  \textit{ Nucl. Instrum. Meth. A} \textbf{ 506} (2003) 250,
  \href{http://dx.doi.org/10.1016/S0168-9002(03)01368-8}{\doi{10.1016/S0168-9002(03)01368-8}}.

\bibitem{CMS:2017yfk}
\hrefCMSnoop {}{{CMS Collaboration}, ``{Particle-flow reconstruction and global
  event description with the CMS detector}'',} \textit{ JINST} \textbf{ 12}
  (2017) P10003,
  \href{http://dx.doi.org/10.1088/1748-0221/12/10/P10003}{\doi{10.1088/1748-0221/12/10/P10003}},
  \href{http://www.arXiv.org/abs/1706.04965}{\texttt{arXiv:1706.04965}}.

\bibitem{CMS:2022uhn}
\hrefCMSnoop {}{{CMS Collaboration}, ``{Measurements of the Higgs boson
  production cross section and couplings in the W boson pair decay channel in
  proton-proton collisions at $\sqrt{s}=13\,\text {Te\hspace{-.08em}V} $}'',}
  \textit{ Eur. Phys. J. C} \textbf{ 83} (2023) 667,
  \href{http://dx.doi.org/10.1140/epjc/s10052-023-11632-6}{\doi{10.1140/epjc/s10052-023-11632-6}},
  \href{http://www.arXiv.org/abs/2206.09466}{\texttt{arXiv:2206.09466}}.

\bibitem{CMS:2015xaf}
\hrefCMSnoop {}{{CMS Collaboration}, ``{Performance of electron reconstruction
  and selection with the CMS detector in proton-proton collisions at
  \ensuremath{\sqrt{s}} = 8 TeV}'',} \textit{ JINST} \textbf{ 10} (2015)
  P06005,
  \href{http://dx.doi.org/10.1088/1748-0221/10/06/P06005}{\doi{10.1088/1748-0221/10/06/P06005}},
  \href{http://www.arXiv.org/abs/1502.02701}{\texttt{arXiv:1502.02701}}.

\bibitem{CMS:2023dsu}
\hrefCMSnoop {}{{CMS Collaboration}, ``{Muon identification using multivariate
  techniques in the CMS experiment in proton-proton collisions at
  sqrt(s)~=~13~TeV}'',} \textit{ JINST} \textbf{ 19} (2024) P02031,
  \href{http://dx.doi.org/10.1088/1748-0221/19/02/P02031}{\doi{10.1088/1748-0221/19/02/P02031}},
  \href{http://www.arXiv.org/abs/2310.03844}{\texttt{arXiv:2310.03844}}.

\bibitem{Cacciari:2008gp}
\hrefCMSnoop {}{M.~Cacciari, G.~P. Salam, and G.~Soyez, ``{The anti-\kt jet
  clustering algorithm}'',} \textit{ JHEP} \textbf{ 04} (2008) 063,
  \href{http://dx.doi.org/10.1088/1126-6708/2008/04/063}{\doi{10.1088/1126-6708/2008/04/063}},
  \href{http://www.arXiv.org/abs/0802.1189}{\texttt{arXiv:0802.1189}}.

\bibitem{Cacciari:2011ma}
\hrefCMSnoop {}{M.~Cacciari, G.~P. Salam, and G.~Soyez, ``{FastJet user
  manual}'',} \textit{ Eur. Phys. J. C} \textbf{ 72} (2012) 1896,
  \href{http://dx.doi.org/10.1140/epjc/s10052-012-1896-2}{\doi{10.1140/epjc/s10052-012-1896-2}},
  \href{http://www.arXiv.org/abs/1111.6097}{\texttt{arXiv:1111.6097}}.

\bibitem{Cacciari:2007fd}
\hrefCMSnoop {}{M.~Cacciari and G.~P. Salam, ``Pileup subtraction using jet
  areas'',} \textit{ Phys. Lett. B} \textbf{ 659} (2008) 119,
  \href{http://dx.doi.org/10.1016/j.physletb.2007.09.077}{\doi{10.1016/j.physletb.2007.09.077}},
  \href{http://www.arXiv.org/abs/0707.1378}{\texttt{arXiv:0707.1378}}.

\bibitem{Bertolini:2014bba}
\hrefCMSnoop {}{D.~Bertolini, P.~Harris, M.~Low, and N.~Tran, ``Pileup per
  particle identification'',} \textit{ JHEP} \textbf{ 10} (2014) 059,
  \href{http://dx.doi.org/10.1007/JHEP10(2014)059}{\doi{10.1007/JHEP10(2014)059}},
\href{http://www.arXiv.org/abs/1407.6013}{\texttt{arXiv:1407.6013}}.
%%CITATION = ARXIV:1407.6013;%%.

\bibitem{Dasgupta:2013ihk}
\hrefCMSnoop {}{M.~Dasgupta, A.~Fregoso, S.~Marzani, and G.~P. Salam, ``Towards
  an understanding of jet substructure'',} \textit{ JHEP} \textbf{ 09} (2013)
  029,
  \href{http://dx.doi.org/10.1007/JHEP09(2013)029}{\doi{10.1007/JHEP09(2013)029}},
\href{http://www.arXiv.org/abs/1307.0007}{\texttt{arXiv:1307.0007}}.
%%CITATION = ARXIV:1307.0007;%%.

\bibitem{Larkoski:2014wba}
\hrefCMSnoop {}{A.~J. Larkoski, S.~Marzani, G.~Soyez, and J.~Thaler, ``Soft
  drop'',} \textit{ JHEP} \textbf{ 05} (2014) 146,
  \href{http://dx.doi.org/10.1007/JHEP05(2014)146}{\doi{10.1007/JHEP05(2014)146}},
\href{http://www.arXiv.org/abs/1402.2657}{\texttt{arXiv:1402.2657}}.
%%CITATION = ARXIV:1402.2657;%%.

\bibitem{Thaler:2010tr}
\hrefCMSnoop {}{J.~Thaler and K.~Van~Tilburg, ``Identifying boosted objects
  with {$N$}-subjettiness'',} \textit{ JHEP} \textbf{ 03} (2011) 015,
  \href{http://dx.doi.org/10.1007/JHEP03(2011)015}{\doi{10.1007/JHEP03(2011)015}},
\href{http://www.arXiv.org/abs/1011.2268}{\texttt{arXiv:1011.2268}}.
%%CITATION = ARXIV:1011.2268;%%.

\bibitem{CMS:2019kuk}
\hrefCMSnoop {}{{CMS Collaboration}, ``{Pileup mitigation at CMS in 13 TeV
  data}'',} \textit{ JINST} \textbf{ 15} (2020) P09018,
  \href{http://dx.doi.org/10.1088/1748-0221/15/09/P09018}{\doi{10.1088/1748-0221/15/09/P09018}},
  \href{http://www.arXiv.org/abs/2003.00503}{\texttt{arXiv:2003.00503}}.

\bibitem{CMS:2017wtu}
\hrefCMSnoop {}{{CMS Collaboration}, ``{Identification of heavy-flavour jets
  with the CMS detector in pp collisions at 13 TeV}'',} \textit{ JINST}
  \textbf{ 13} (2018) P05011,
  \href{http://dx.doi.org/10.1088/1748-0221/13/05/P05011}{\doi{10.1088/1748-0221/13/05/P05011}},
  \href{http://www.arXiv.org/abs/1712.07158}{\texttt{arXiv:1712.07158}}.

\bibitem{CMS:2018lgn}
\hrefCMSnoop {}{{CMS Collaboration}, ``{Measurement of the single top quark and
  antiquark production cross sections in the $t$ channel and their ratio in
  proton-proton collisions at $\sqrt{s}=$ 13 TeV}'',} \textit{ Phys. Lett. B}
  \textbf{ 800} (2020) 135042,
  \href{http://dx.doi.org/10.1016/j.physletb.2019.135042}{\doi{10.1016/j.physletb.2019.135042}},
  \href{http://www.arXiv.org/abs/1812.10514}{\texttt{arXiv:1812.10514}}.

\bibitem{ParticleDataGroup:2024cfk}
\hrefCMSnoop {}{{Particle Data Group}, S.~Navas { et~al.}, ``Review of particle
  physics'',} \textit{ Phys. Rev. D} \textbf{ 110} (2024) 030001,
  \href{http://dx.doi.org/10.1103/PhysRevD.110.030001}{\doi{10.1103/PhysRevD.110.030001}}.

\bibitem{Lindert:2017olm}
\hrefCMSnoop {}{J.~M. Lindert { et~al.}, ``{Precise predictions for $V+$jets
  dark matter backgrounds}'',} \textit{ Eur. Phys. J. C} \textbf{ 77} (2017)
  829,
  \href{http://dx.doi.org/10.1140/epjc/s10052-017-5389-1}{\doi{10.1140/epjc/s10052-017-5389-1}},
  \href{http://www.arXiv.org/abs/1705.04664}{\texttt{arXiv:1705.04664}}.

\bibitem{CMS:2018zzl}
\hrefCMSnoop {}{{CMS Collaboration}, ``{Measurements of properties of the Higgs
  boson decaying to a W boson pair in pp collisions at $\sqrt{s}=$ 13 TeV}'',}
  \textit{ Phys. Lett. B} \textbf{ 791} (2019) 96,
  \href{http://dx.doi.org/10.1016/j.physletb.2018.12.073}{\doi{10.1016/j.physletb.2018.12.073}},
  \href{http://www.arXiv.org/abs/1806.05246}{\texttt{arXiv:1806.05246}}.

\bibitem{keras}
\hrefCMSnoop {}{F.~Chollet { et~al.}, ``Keras'',} 2015.
  https://github.com/fchollet/keras.

\bibitem{tensorflow2015-whitepaper}
\hrefCMSnoop {}{M.~Abadi { et~al.}, ``\textsc{TensorFlow}: Large-scale machine
  learning on heterogeneous systems'',} 2015.
  \href{http://www.arXiv.org/abs/1603.04467}{\texttt{arXiv:1603.04467}}.

\bibitem{kingma2017adam}
\hrefCMSnoop {}{D.~P. Kingma and J.~Ba, ``Adam: A method for stochastic
  optimization'',} \textit{ Proceedings of the 3rd International Conference for
  Learning Representations, San Diego, 2015} (2017)
  \href{http://dx.doi.org/10.48550/arXiv.1412.6980}{\doi{10.48550/arXiv.1412.6980}},
  \href{http://www.arXiv.org/abs/1412.6980}{\texttt{arXiv:1412.6980}}.

\bibitem{agarap2018deep}
\hrefCMSnoop {}{A.~F. Agarap, ``Deep learning using rectified linear units
  ({ReLU})'',} 2018.
  \href{http://www.arXiv.org/abs/1803.08375}{\texttt{arXiv:1803.08375}}.

\bibitem{Goodfellow-et-al-2016}
I.~Goodfellow, Y.~Bengio, and A.~Courville, ``Deep learning''.
\newblock MIT Press, 2016.
\newblock \url{http://www.deeplearningbook.org}. ISBN~0262035618.

\bibitem{bayes-opt}
\href {https://github.com/fmfn/BayesianOptimization}{F.~Nogueira, ``{Bayesian
  optimization}: Open source constrained global optimization tool for
  {Python}'',} 2014.
\newblock \url {https://github.com/fmfn/BayesianOptimization}.

\bibitem{shap}
\hrefCMSnoop {}{S.~M. Lundberg and S.-I. Lee, ``A unified approach to
  interpreting model predictions'',} in \textit{ Proceedings of the 31st
  International Conference on Neural Information Processing Systems}, NIPS'17,
  p.~4768.
\newblock Curran Associates Inc., Red Hook, NY, USA, 2017.
\newblock
  \href{http://www.arXiv.org/abs/1705.07874}{\texttt{arXiv:1705.07874}}.

\bibitem{Rainwater:1996ud}
\hrefCMSnoop {}{D.~L. Rainwater, R.~Szalapski, and D.~Zeppenfeld, ``{Probing
  color singlet exchange in $Z$ + 2-jet events at the CERN LHC}'',} \textit{
  Phys. Rev. D} \textbf{ 54} (1996) 6680,
  \href{http://dx.doi.org/10.1103/PhysRevD.54.6680}{\doi{10.1103/PhysRevD.54.6680}},
  \href{http://www.arXiv.org/abs/hep-ph/9605444}{\texttt{arXiv:hep-ph/9605444}}.

\bibitem{CMS:2013kfa}
\href {https://cds.cern.ch/record/1599732}{{CMS Collaboration}, ``Performance
  of quark/gluon discrimination in {$8\TeV$} {$\Pp\Pp$} data'',} CMS Physics
  Analysis Summary CMS-PAS-JME-13-002, 2013.

\bibitem{CMS-DP-2016-070}
\href {https://cds.cern.ch/record/2234117}{{CMS Collaboration}, ``Performance
  of quark/gluon discrimination in {$13\TeV$} data'',} CMS Detector Performance
  Summary CMS-DP-2016-070, 2016.

\bibitem{wilks}
\hrefCMSnoop {}{G.~Cowan, K.~Cranmer, E.~Gross, and O.~Vitells, ``{Asymptotic
  formulae for likelihood-based tests of new physics}'',} \textit{ Eur. Phys.
  J. C} \textbf{ 71} (2011) 1554,
  \href{http://dx.doi.org/10.1140/epjc/s10052-011-1554-0}{\doi{10.1140/epjc/s10052-011-1554-0}},
  \href{http://www.arXiv.org/abs/1007.1727}{\texttt{arXiv:1007.1727}}.
  [Erratum: Eur. Phys. J. C 73, 2501 (2013)].

\bibitem{CMS:2024onh}
\hrefCMSnoop {}{{CMS Collaboration}, ``{The CMS statistical analysis and
  combination tool: COMBINE}'',} \textit{ Comput. Softw. Big Sci.} \textbf{ 8}
  (2024) 19,
  \href{http://dx.doi.org/10.1007/s41781-024-00121-4}{\doi{10.1007/s41781-024-00121-4}},
  \href{http://www.arXiv.org/abs/2404.06614}{\texttt{arXiv:2404.06614}}.

\bibitem{Verkerke:2003ir}
\href {http://inspirehep.net/record/634021}{W.~Verkerke and D.~P. Kirkby,
  ``{The RooFit toolkit for data modeling}'',} in \textit{ {Proceedings of the
  13th International Conference for Computing in High-Energy and Nuclear
  Physics (CHEP03)}}.
\newblock 2003.
\newblock
  \href{http://www.arXiv.org/abs/physics/0306116}{\texttt{arXiv:physics/0306116}}.
\newblock
{[eConf C0303241, MOLT007]}.
%%CITATION = PHYSICS/0306116;%%.

\bibitem{Moneta:2010pm}
L.~Moneta\hrefCMSnoop {}{ { et~al.}, ``{The RooStats project}'',} \textit{ PoS}
  \textbf{ ACAT2010} (2010) 057,
  \href{http://dx.doi.org/10.22323/1.093.0057}{\doi{10.22323/1.093.0057}},
  \href{http://www.arXiv.org/abs/1009.1003}{\texttt{arXiv:1009.1003}}.

\bibitem{lumi16}
\hrefCMSnoop {}{{CMS Collaboration}, ``Precision luminosity measurement in
  proton-proton collisions at $\sqrt{s}={13\TeV}$ in 2015 and 2016 at {CMS}'',}
  \textit{ Eur. Phys. J. C} \textbf{ 81} (2021) 800,
  \href{http://dx.doi.org/10.1140/epjc/s10052-021-09538-2}{\doi{10.1140/epjc/s10052-021-09538-2}},
  \href{http://www.arXiv.org/abs/2104.01927}{\texttt{arXiv:2104.01927}}.

\bibitem{lumi17}
\href {http://cds.cern.ch/record/2621960}{{CMS Collaboration}, ``{CMS}
  luminosity measurement for the 2017 data-taking period at
  $\sqrt{s}={13\TeV}$'',} CMS Physics Analysis Summary CMS-PAS-LUM-17-004,
  2018.

\bibitem{lumi18}
\href {http://cds.cern.ch/record/2676164}{{CMS Collaboration}, ``{CMS}
  luminosity measurement for the 2018 data-taking period at
  $\sqrt{s}={13\TeV}$'',} CMS Physics Analysis Summary CMS-PAS-LUM-18-002,
  2019.

\bibitem{CMS:2018mlc}
\hrefCMSnoop {}{{CMS Collaboration}, ``{Measurement of the inelastic
  proton-proton cross section at $ \sqrt{s}=13 $ TeV}'',} \textit{ JHEP}
  \textbf{ 07} (2018) 161,
  \href{http://dx.doi.org/10.1007/JHEP07(2018)161}{\doi{10.1007/JHEP07(2018)161}},
  \href{http://www.arXiv.org/abs/1802.02613}{\texttt{arXiv:1802.02613}}.

\bibitem{CMS:2024ppo}
\hrefCMSnoop {}{{CMS Collaboration}, ``{Performance of the CMS electromagnetic
  calorimeter in pp collisions at $\sqrt{s}=$ 13 TeV}'',} \textit{ JINST}
  \textbf{ 19} (2024) P09004,
  \href{http://dx.doi.org/10.1088/1748-0221/19/09/P09004}{\doi{10.1088/1748-0221/19/09/P09004}},
  \href{http://www.arXiv.org/abs/2403.15518}{\texttt{arXiv:2403.15518}}.

\bibitem{Khachatryan:2016kdb}
\hrefCMSnoop {}{{CMS Collaboration}, ``{Jet energy scale and resolution in the
  CMS experiment in pp collisions at 8 TeV}'',} \textit{ JINST} \textbf{ 12}
  (2017) P02014,
  \href{http://dx.doi.org/10.1088/1748-0221/12/02/P02014}{\doi{10.1088/1748-0221/12/02/P02014}},
  \href{http://www.arXiv.org/abs/1607.03663}{\texttt{arXiv:1607.03663}}.

\bibitem{CMS:2014rsx}
\hrefCMSnoop {}{{CMS Collaboration}, ``{Identification techniques for highly
  boosted W bosons that decay into hadrons}'',} \textit{ JHEP} \textbf{ 12}
  (2014) 017,
  \href{http://dx.doi.org/10.1007/JHEP12(2014)017}{\doi{10.1007/JHEP12(2014)017}},
  \href{http://www.arXiv.org/abs/1410.4227}{\texttt{arXiv:1410.4227}}.

\bibitem{Kalogeropoulos:2018cke}
\hrefCMSnoop {}{A.~Kalogeropoulos and J.~Alwall, ``{The SysCalc code: A tool to
  derive theoretical systematic uncertainties}'',} 2018.
  \href{http://www.arXiv.org/abs/1801.08401}{\texttt{arXiv:1801.08401}}.

\bibitem{Rojo:2016ymp}
\hrefCMSnoop {}{J.~Rojo, ``{PDF4LHC recommendations for Run II}'',} \textit{
  PoS} \textbf{ DIS2016} (2016) 018,
  \href{http://dx.doi.org/10.22323/1.265.0018}{\doi{10.22323/1.265.0018}},
  \href{http://www.arXiv.org/abs/1606.08243}{\texttt{arXiv:1606.08243}}.

\bibitem{BB}
\hrefCMSnoop {}{J.~S. Conway, ``{Incorporating nuisance parameters in
  likelihoods for multisource spectra}'',} \textit{ {PHYSTAT 2011}} (2011) 115,
  \href{http://dx.doi.org/10.5170/CERN-2011-006.115}{\doi{10.5170/CERN-2011-006.115}},
  \href{http://www.arXiv.org/abs/1103.0354}{\texttt{arXiv:1103.0354}}.

\bibitem{Bernlochner:2022oiw}
\hrefCMSnoop {}{F.~U. Bernlochner, D.~C. Fry, S.~B. Menary, and E.~Persson,
  ``{Cover your bases: asymptotic distributions of the profile likelihood ratio
  when constraining effective field theories in high-energy physics}'',}
  \textit{ SciPost Phys. Core} \textbf{ 6} (2023) 013,
  \href{http://dx.doi.org/10.21468/SciPostPhysCore.6.1.013}{\doi{10.21468/SciPostPhysCore.6.1.013}},
  \href{http://www.arXiv.org/abs/2207.01350}{\texttt{arXiv:2207.01350}}.

\bibitem{ATLAS:2022nru}
\hrefCMSnoop {}{{ATLAS Collaboration}, ``{Measurement of electroweak $
  Z\left(\nu \overline{\nu}\right)\gamma jj $ production and limits on
  anomalous quartic gauge couplings in pp collisions at $ \sqrt{s} $ = 13 TeV
  with the ATLAS detector}'',} \textit{ JHEP} \textbf{ 06} (2023) 082,
  \href{http://dx.doi.org/10.1007/JHEP06(2023)082}{\doi{10.1007/JHEP06(2023)082}},
  \href{http://www.arXiv.org/abs/2208.12741}{\texttt{arXiv:2208.12741}}.

\end{thebibliography}\endgroup
